\documentclass[%
 a4paper
]{article}
\pdfoutput=1
\usepackage{jheppub}
\usepackage{graphicx}
\usepackage{dcolumn}
\usepackage{bm}
\usepackage{slashed}
\usepackage{color}
\usepackage{physics}
\usepackage{comment}
\usepackage[normalem]{ulem}
\usepackage{hyperref}
\usepackage{url}
\usepackage{feynmp-auto}

\renewcommand{\vev}[1]{\langle #1 \rangle}
\newcommand{\MPl}{M_{\rm Pl}}

\begin{document}



\title{
Accidental Peccei-Quinn Symmetry from \\ Chiral Gauge Symmetry and Mirror QCD 
}

\author[a]{Hajime Fukuda}
\emailAdd{hfukuda@hep-th.phys.s.u-tokyo.ac.jp}
\affiliation[a]{Department of Physics, The University of Tokyo, Tokyo 113-0033, Japan}
\author[b,c]{Keisuke Harigaya}
\emailAdd{kharigaya@uchicago.edu}
\affiliation[b]{Enrico Fermi Institute, Kavli Institute for Cosmological Physics, Leinweber Institute for Theoretical Physics, Department of Physics, The University of Chicago, Chicago, IL 60637, USA}
\affiliation[c]{Kavli Institute for the Physics and Mathematics of the Universe (WPI),
The University of Tokyo Institutes for Advanced Study,
The University of Tokyo, Kashiwa, Chiba 277-8583, Japan}

\date{\today}

\abstract{
We present a solution to the strong CP problem in which a simple chiral \(U(1)\) gauge symmetry gives rise to an accidental Peccei-Quinn symmetry that is both explicitly and spontaneously broken by mirror QCD dynamics, yielding a framework without massless fermions or a light QCD axion. The model contains no stable domain walls or colored relics, and it accommodates a sufficiently high reheating temperature to account for the baryon asymmetry of the Universe via leptogenesis. Metastable domain walls and first-order mirror QCD phase transition generate a stochastic background of primordial gravitational waves. Additionally, one of the pseudo-Nambu-Goldstone bosons serves as a viable WIMP dark matter candidate.  The gauge boson associated with the chiral \(U(1)\) gauge symmetry, which kinetically mixes with the Standard Model hypercharge gauge boson, provides a vector portal connecting dark matter to the Standard Model and plays a central role in the dark-matter phenomenology. Colored pseudo-Nambu-Goldstone bosons can be probed at the LHC through searches for dijet resonances, jets plus missing energy, multijet events with leptons, and displaced vertices.
}

\maketitle
\newpage

\section{Introduction}

The strong CP problem is one of the most puzzling issues in particle physics. The QCD Lagrangian allows a CP-violating term, but its coefficient, $\theta$, is experimentally constrained to be extremely small, $|\theta| \lesssim 10^{-10}$\,\cite{Baker:2006ts,Pospelov:1999ha}. This raises the question of why $\theta$ is so suppressed, which is referred to as the strong CP problem. One of the most well-known solutions is the Peccei--Quinn (PQ) mechanism~\cite{Peccei:1977hh,Peccei:1977ur}, in which a new global symmetry is spontaneously broken, giving rise to a pseudo-Nambu--Goldstone boson (NGB), the axion~\cite{Weinberg:1977ma,Wilczek:1977pj}. The axion dynamically relaxes $\theta$ to zero, thereby solving the strong CP problem.

However, the PQ mechanism itself faces conceptual challenges. There is no fundamental reason for the existence of the PQ ``symmetry,'' which is not an exact symmetry due to the QCD anomaly. At best, it is an accidental symmetry that is explicitly broken by higher-dimensional operators~\cite{Holman:1992us,Barr:1992qq,Kamionkowski:1992mf,Dine:1992vx}, similar to baryon number in the Standard Model (SM). Moreover, it is widely believed that exact global symmetries do not exist in quantum gravity\,\cite{Kallosh:1995hi,Arkani-Hamed:2006emk,Banks:2010zn,Harlow:2018tng}, so the PQ symmetry is also expected to be violated by Planck-suppressed operators. Such explicit breaking generically induces an additional contribution to the axion potential, shifting its minimum away from $\theta=0$. Since the QCD-induced axion potential is suppressed by the ratio of the QCD scale to the PQ-breaking scale, even tiny explicit breaking at high energies can lead to an unacceptably large shift in $\theta$. Consequently, maintaining $|\theta| \lesssim 10^{-10}$ requires extreme suppression of PQ-violating operators, leading to the so-called axion quality problem.

One approach to the axion quality problem is to introduce  (possibly discrete) gauge symmetry such that the PQ symmetry arises accidentally from the gauge structure and matter content~\cite{Georgi:1981pu,Barr:1992qq,Lazarides:1985bj,Randall:1992ut,Dias:2002hz,Dias:2002gg,Babu:2002ic,Choi:2009jt,Carpenter:2009zs,Harigaya:2013vja,Harigaya:2015soa,Fukuda:2017ylt,DiLuzio:2017tjx,Duerr:2017amf,Ibe:2018hir,Lillard:2018fdt,Fukuda:2018oco,Nakai:2021nyf,Contino:2021ayn,Choi:2022fha,Gherghetta:2025kff,Agrawal:2025mke,Babu:2026yqp}. In such constructions, suitable charge assignments can forbid lower-dimensional PQ-violating operators. The quality problem can be further alleviated if the PQ symmetry is broken dynamically by strong dynamics, since PQ-violating operators typically arise at higher dimensions when the fundamental degrees of freedom are fermionic rather than scalar. Moreover, dynamical symmetry breaking avoids a hierarchy between the PQ-breaking scale and the Planck scale~\cite{Choi:1985cb}. However, given the stringent bound on $\theta$\,\cite{Baker:2006ts}, one must typically forbid operators up to dimension $\gtrsim 10$, which often necessitates intricate charge assignments and/or multiple gauge symmetries.

Another approach is to make the axion heavier, thereby reducing the sensitivity of its potential to explicit breaking. This can be achieved by introducing a hidden strongly coupled gauge sector at a high scale that generates an additional large contribution to the axion potential. For the axion to cancel the QCD $\theta$ term, the $\theta$ parameters in the visible and hidden sectors must be aligned. This alignment can be enforced by a discrete symmetry relating the two sectors~\cite{Rubakov:1997vp,Berezhiani:2000gh,Hook:2014cda,Fukuda:2015ana,Dunsky:2023ucb}.%
\footnote{
Alternatively, one may embed $SU(3)_c$ into a larger gauge group broken to $SU(3)_c \times G$ and generate the axion potential from the dynamics of $G$~\cite{Gherghetta:2016fhp,Gaillard:2018xgk,Valenti:2022tsc,Bedi:2024kxe}, or enhance the $SU(3)_c$ coupling at high energies~\cite{Flynn:1987rs,Agrawal:2017ksf,Csaki:2019vte,Gherghetta:2020keg,Kitano:2021fdl,Aoki:2024usv}.}
Since the physical $\theta$ angle depends on the phases of quark masses, the discrete symmetry must also relate the Yukawa structures of the two sectors. This requires the introduction of a mirror copy of the Standard Model, related by a $Z_2$ symmetry. However, in such setups, the origin of the PQ symmetry remains unexplained unlike in the case of an accidental symmetry, and the axion quality problem is not fully resolved.

In this paper, we combine these approaches to construct a solution to the strong CP problem without introducing fine-tuning. First, we revisit the model proposed in~\cite{Hook:2014cda}. In this framework, the PQ symmetry is spontaneously broken by mirror QCD dynamics, eliminating a hierarchy problem between the PQ-breaking scale and the Planck scale. However, in the original construction, the PQ symmetry is imposed as an anomalous symmetry, and its origin and quality remain unexplained. We show that the model admits an anomaly-free $Z_3$ symmetry, which, if imposed as an exact discrete gauge symmetry, leads to an accidental PQ symmetry and resolves the quality problem. Nevertheless, this setup suffers from an unavoidable domain wall problem, requiring a low reheating temperature, and predicts a stable colored particle whose cosmological viability depends on assumptions about early-universe dynamics and bound-state formation.

We then propose an alternative mirror axion model in which the PQ symmetry arises from a chiral $U(1)$ gauge symmetry. In this construction, both the domain wall problem and stable colored relics are avoided, and the reheating temperature can be sufficiently high to allow for leptogenesis. The axion acquires a large mass from mirror QCD dynamics, rendering $\theta=0$ robust against explicit PQ breaking from higher-dimensional operators. A simple chiral $U(1)$ charge assignment, analogous to those in~\cite{Harigaya:2016rwr,Co:2016akw,Contino:2020god,Ibe:2021gil}, is sufficient to solve the strong CP problem.
In addition, we investigate the phenomenology of the model. We find that 
the model predicts a dark matter candidate with a vector portal testable by indirect detection experiments, as well as signatures at colliders (jets plus missing energy, multi jets plus leptons, displaced vertices, and dijet resonances), electric dipole moment measurements, and primordial gravitational waves.

This paper is organized as follows. In Sec.~\ref{sec:Z3}, we review the model of~\cite{Hook:2014cda} and highlight the role of an exact $Z_3$ symmetry in realizing an accidental PQ symmetry. In Sec.~\ref{sec:U1}, we present the $U(1)$ gauge model and demonstrate the absence of stable domain walls and colored relics, followed by a discussion of its phenomenology. We summarize our results in Sec.~\ref{sec:summary}.

\section{Accidental PQ from $Z_3$}
\label{sec:Z3}

In this section, we revisit the model proposed in~\cite{Hook:2014cda} from the perspective of the origins of the PQ symmetry.
In the original work, the PQ symmetry is assumed to exist without an underlying explanation. We point out that, in the absence of lower-dimensional PQ-breaking operators, the model admits an anomaly-free $Z_3$ symmetry. If this symmetry is imposed as an exact discrete gauge symmetry, the PQ symmetry can arise accidentally, thereby addressing the axion quality problem. 
However, we find that this construction inevitably leads to a domain wall problem if the $Z_3$ symmetry is exact. In addition, the model predicts a stable colored particle, and we discuss the possibility that its bound state contributes to the dark matter abundance.

\subsection{Setup}

The SM is extended with its mirror copy, SM$'$, and a $Z_2$ symmetry is introduced to exchange the SM and SM$'$. The $Z_2$ symmetry ensures that the $\theta$ term of the SM and SM$'$ are identical. In addition, a pair of massless fermions $\psi$ and $\bar{\psi}$ with the $SU(3)_c\times SU(3)_c'$ charges shown in Table~\ref{tab:charge1} are introduced. The $Z_2$ symmetry acts on each component of $\psi$ and $\bar{\psi}$ as $\psi_{i j} \leftrightarrow \psi_{j i}$, where the first and second indices are for $SU(3)_c$ and $SU(3)_c'$, respectively. 

With these fermions, the $\theta$ term of the SM$'$ can be removed by the chiral rotation of $\psi$ and $\bar{\psi}$. This can be understood as a result of the $U(1)_{\rm PQ}$ symmetry in Table~\ref{tab:charge1}. Here, we assume that the $U(1)_{\rm PQ}$ symmetry is violated only by the Adler-Bell-Jackiw (ABJ) anomaly of $SU(3)_c$ and $SU(3)_c'$. Because of the $Z_2$ symmetry, the $\theta$ term of the SM also vanishes in this basis. 

We assume that the $Z_2$ symmetry is spontaneously broken by some mechanism~\cite{Blinov:2016kte,Hall:2018let}, so that the mirror electroweak (EW) scale $v'$ is much larger than the EW scale $v$. The details of the mechanism of the $Z_2$ breaking is not important for the following discussion, and we do not specify it, but we require that the $Z_2$ breaking does not introduce large CP-violating phases and the $\theta$ term is not regenerated by the $Z_2$ breaking.

\begin{figure}
    \centering
    \includegraphics[width=0.7\linewidth]{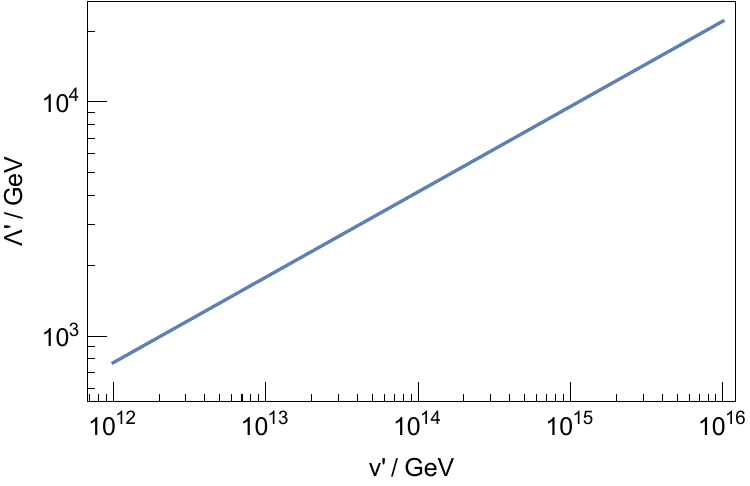}
    \caption{The mirror QCD scale $\Lambda'$ as a function of the mirror electroweak scale $v'$ in the model in Sec.~\ref{sec:Z3}.}
    \label{fig:vpL0}
\end{figure}

With large $v'$, the mirror EW symmetry is broken at a high scale, and mirror quarks are heavier than the SM quark. This changes the running of the gauge coupling constant and the mirror QCD scale $\Lambda'$ is much larger than the QCD scale $\Lambda$. Let us assume that $y_u v' > \Lambda'$, where $y_u$ is the up-quark Yukawa coupling, for the moment. We will check the consistency of this assumption later. In this case, all the mirror quarks except for $\psi$ and $\bar{\psi}$ are heavier than the mirror QCD scale. In Fig.~\ref{fig:vpL0}, we show $\Lambda'$ as a function of $v'$. Here $\Lambda'$ is defined as an energy scale at which the $SU(3)_c'$ gauge coupling computed at the two-loop level diverges.

The mirror QCD dynamics is that of a three-flavor QCD, where the approximate chiral symmetry $SU(3)_L\times SU(3)_R$ is spontaneously broken down to the diagonal subgroup, which is $SU(3)_c$. The symmetry breaking gives rise to eight NGBs, which are octets of $SU(3)_c$. Because the axial part of the chiral symmetry is explicitly broken by $SU(3)_c$, the pseudo NGBs obtain a mass $ m_O\sim 0.8 g_3 \Lambda'$\,\cite{Contino:2020god}. The QCD axion, which corresponds to the $\eta'$ meson of the mirror QCD, obtains a mass $\sim 4 \Lambda'$ by the mirror QCD dynamics.%
\footnote{
    Since there are no other contributions to the $\eta'$ potential, it is expected not to have any non-trivial branch structures\,\cite{Witten:1980sp}.
}
The mirror dynamics also produces baryons made from $\psi$ and $\bar{\psi}$. The cosmology of the baryons is discussed in Sec.~\ref{sec:baryon}. After integrating out the pseudo NGBs and baryons, there remain no new particles beyond the SM.

The existence of light colored particles, the octet NGBs, gives the lower bound on $\Lambda'$ from collider searches.
The octet NGBs can be pair-produced at the LHC and decay into a pair of gluons by the Wess-Zumino-Witten term. The bound from this process is $m_O \gtrsim 800$ GeV\,\cite{ATLAS:2017jnp}, which translates into $v' > 10^{12}$ GeV. With this bound, we find that $y_u v' > \Lambda'$ is satisfied, so the assumption we made is consistent. 

On the other hand, with large $v'$, the $\theta$ term of the SM QCD might not be zero anymore. Indeed, from the viewpoint of symmetry, the following interaction
\begin{equation}
   \frac{1}{32\pi^2}\frac{|H|^2}{M_H^2} G\tilde{G} + \frac{1}{32\pi^2}\frac{|H'|^2}{M_H^2} G'\tilde{G'}
\end{equation}
cannot be forbidden, where $M_H$ is a cutoff scale.%
\footnote{
If CP is preserved around the cutoff scale and is spontaneously broken at a lower energy scale, this interaction can be suppressed.
}
Because $v'\gg v$, the $\theta$ terms of $SU(3)_c$ is no longer equal to that of $SU(3)_c'$, resulting in a non-zero strong CP phase,
\begin{equation}
\label{eq:dtheta}
    \Delta \theta = 10^{-10} \left(\frac{v'}{10^{13}~{\rm GeV}}\right)^2\left(\frac{10^{18}~{\rm GeV}}{M_H}\right)^2.
\end{equation}
For $M_H$ around the reduced Planck scale, $v' \lesssim $ few $10^{13}$ GeV is consistent with the upper bound on the strong CP phase.

\begin{table}
    \centering
    \begin{tabular}{c|ccc|c}
         & $SU(3)_c$ & $SU(3)_c'$ & $Z_{3}$ & $U(1)_{\rm PQ}$ \\ \hline
        $\psi$ & ${\bf 3}$ & ${\bf 3}$ & $1$ &  $1/2$ \\
        $\bar{\psi}$ & $\overline{{\bf 3}}$ & $\overline{{\bf 3}}$ & $0$  & $1/2$ \\
    \end{tabular}
    \caption{The quantum numbers of fermions $\psi$ and $\bar{\psi}$. $Z_{3}$ does not have $SU(3)_c$ or $SU(3)_c'$ anomaly and may be an exact gauge symmetry. $U(1)_{\rm PQ}$ is an accidental symmetry arising from $Z_{3}$.}
    \label{tab:charge1}
\end{table}

At this point, the PQ symmetry, $U(1)_{\rm PQ}$, is an anomalous symmetry that is explicitly broken by the ABJ anomaly of $SU(3)_c$ and $SU(3)_c'$ and we do not have any reason to expect its existence or quality; it may well be broken even by a dimension-3 operator $\psi \bar{\psi}$.
However, we point out that the model admits an anomaly-free $Z_3$ symmetry shown in Table~\ref{tab:charge1}. By imposing $Z_{3}$ symmetry as an exact gauge symmetry,  the PQ symmetry can arise accidentally and the axion quality problem can be solved. For example, the dimension-3 operator $\psi \bar{\psi}$ is forbidden by the $Z_{3}$ symmetry.

Since we assume that the PQ symmetry is an accidental symmetry, it is explicitly broken by higher-dimensional operators. The most relevant operator which breaks the PQ symmetry and contributes to the axion potential is the dimension-9 operator $(\psi \bar{\psi})^3/M^5$, where $M$ is a cutoff scale.%
\footnote{
    Dimension-8 operators such as $(\psi D_\mu \psi) (\psi \gamma^\mu L)H $ are also allowed by the $Z_3$ symmetry, but they do not contribute to the axion potential. 
}
After removing the $\theta$ term, the coefficient of this operator is in general complex and violates the CP symmetry. The resultant correction to the $\theta$ term is $O(\Lambda'^5/ M^5)$. To solve the strong CP problem requires that
\begin{equation}
    \Lambda' \lesssim 10^{-2} M = 10^{16}~{\rm GeV} \frac{M}{10^{18}~{\rm GeV}}.
\end{equation}
This is consistent with the constraint from the collider search for colored particles, which requires $\Lambda' >$ TeV, and the model can solve the strong CP problem without the quality problem.

\subsection{Stable domain wall}
\label{sec:Z3_DW}

After the chiral symmetry breaking, this model with an exact $Z_3$ symmetry has a domain wall problem. However, the symmetry structure is slightly different from the usual domain wall problem, so we start with a brief review of the symmetry structure of the model, based on the generic analysis in~\cite{Preskill:1992bf}.
As mentioned above, the model can be regarded as a three-flavor QCD with the gauge group $SU(3)_{c'}$, so first let us consider the limit where the $SU(3)_c$ gauge coupling vanishes and the gauged $Z_3$ symmetry is replaced by a global $Z_3$ symmetry. 
We also ignore higher-dimensional operators that break global symmetries for the moment. 
In that case, the symmetry of the model is $G_\text{approx}\equiv SU(3)_L\times SU(3)_R\times U(1)_V$, where $SU(3)_L$ and $SU(3)_R$ are the chiral symmetries acting on $\psi$ and $\bar{\psi}$, respectively, and $U(1)_V$ is the baryon number symmetry. After the chiral symmetry breaking, the symmetry is broken down to $H_\text{approx}\equiv SU(3)_V\times U(1)_V$, where $SU(3)_V$ is the diagonal subgroup of $SU(3)_L\times SU(3)_R$. 

Let us turn on the $SU(3)_c$ gauge coupling. The symmetry structure of the model is as follows. For the continuous part of the symmetry, $G_\text{approx}$ is explicitly broken down to $SU(3)_c\times U(1)_V$ by the $SU(3)_c$ gauge interaction. However, the discrete $Z_3$ symmetry, which is the center of $SU(3)_{L}$, is not broken by the $SU(3)_c$ gauge interaction as we have seen, so the symmetry of the model is $G_\text{exact} \equiv SU(3)_c\times U(1)_V \times Z_{3}$, which is broken down to $H_\text{exact} \equiv SU(3)_c\times U(1)_V$.

With the symmetry breaking pattern $G_\text{exact} \rightarrow H_\text{exact}$, topological defects are formed. Focusing on the exact symmetries, the vacuum manifold is $G_\text{exact}/H_\text{exact} = Z_3$, which supports $Z_3$ domain walls. However, in the limit where the $SU(3)_c$ gauge coupling vanishes, the vacuum manifold is $G_\text{approx}/H_\text{approx} = \left(SU(3)_L\times SU(3)_L\right)/SU(3)_V$, and there are no stable domain walls. Therefore, we expect that {the domain walls appear due to the explicit breaking of the continuous symmetry $G_\text{approx}$ by the $SU(3)_c$ gauge interaction, and the domain wall tension vanishes in the limit where the $SU(3)_c$ gauge coupling vanishes.}

Indeed, we can see this from the explicit form of the chiral condensation. The chiral condensation, $\langle \psi \bar{\psi} \rangle$, is charged under the $Z_3$ symmetry, so there are three vacua related by the $Z_3$ transformation,
\begin{equation}
\label{eq:vacZ3}
    \vev{\psi \bar{\psi}} \sim \Lambda^{'3}{\rm exp}\left( i \frac{2\pi}{3}k\right),~~k = 0,1,2.
\end{equation}
At first sight, it may seem that these three vacua are separated by a potential barrier from the chiral condensation, and the tension of the domain walls is  ${\cal O}(4\pi\Lambda'^3)$. However, this is not the case, since such domain walls do not exist if the $SU(3)_c$ gauge coupling vanishes as the discussion on the symmetry structure above suggests.
In fact, the three vacua are connected by $G_\text{approx}$ transformation. To see this, let the (pseudo-)NGB associated with the chiral symmetry breaking be $O^a$. Under the $G_\text{approx} / H_\text{approx}$ transformation, $O^a$ and the chiral condensation transform non-linearly. In particular, focusing on the axial rotation $\psi \to e^{i \alpha T^8} \psi$ and $\bar{\psi} \to e^{i \alpha T^8} \bar{\psi}$ with $T^8 = {\rm diag}(1,1,-2)/2\sqrt{3}$, the transformations are
\begin{align}
    O^8 \to O^8 + \alpha f_{\pi'}, ~~
    \vev{\psi \bar{\psi}} \to \vev{\psi e^{2i \alpha T^8} \bar{\psi}},
\end{align}
where $f_{\pi'}$ is the decay constant of the NGBs. For $\alpha = \alpha_0 \equiv -2\pi / \sqrt{3}$, $\langle \psi \bar{\psi} \rangle$ is transformed as $\langle \psi \bar{\psi} \rangle \to \langle \psi \bar{\psi} \rangle e^{-2\pi i / 3}$, and thus the continuous transformation, $\alpha = \alpha_0 s$ with $s \in [0,1]$, continuously connects the three vacua in Eq.~\eqref{eq:vacZ3}. This means that the domain wall tension vanishes in the limit where the $SU(3)_c$ gauge coupling vanishes so that $G_\text{approx}$ is an exact symmetry.

In the full model, the $SU(3)_c$ gauge coupling is nonzero, which explicitly breaks the continuous symmetry $G_{\rm approx}$ and leads to the formation of stable domain walls. In the low-energy effective theory, this effect is captured by a potential for the pseudo-NGBs generated by the $SU(3)_c$ gauge interaction, given by
\begin{align}
    V(O^a) \propto \qty|\text{Tr}(U)|^2 \sim -\cos \qty(\frac{\sqrt{3}O^8}{f_{\pi'}}) + \cdots,
\end{align}
where $U = \exp(i O^a T^a / f_{\pi'})$ and we have focused on the potential for $O^8$ for simplicity. With this potential, the $Z_3$ symmetry, $\alpha \to \alpha + \alpha_0$, is still an exact symmetry, but the continuous transformation, $\alpha = \alpha_0 s$ with $s \in [0,1]$, is not a symmetry anymore and the three vacua are separated by a potential barrier from $V(O^a)$. 

Let us estimate the domain wall tension. 
As long as the $SU(3)_c$ gauge coupling is perturbative, the potential barrier from $V(O^a)$ is smaller than that from the chiral condensation, so the domain wall tension is solely determined by the potential barrier from $V(O^a)$.%
\footnote{
This means that $SU(3)_c$ is broken to $SU(2)\times U(1)$ inside the domain walls.
}
The height of the potential barrier is $\sim m_O^2 f_{\pi'}^2$. On the other hand, the thickness of the domain wall is $\sim 1/m_O$, so the domain wall tension $\sigma$ is $\sim m_O f_{\pi'}^2$. More precisely, we adopt
\begin{equation}
    \sigma \simeq 9 m_O \left(\frac{f_{\pi'}}{\sqrt{3}}\right)^2 \simeq 1.7 g_3 \Lambda^{'3},
\end{equation}
where the factor of $9$ is taken from the estimation for the QCD axion domain wall~\cite{Sikivie:2006ni},
and in the second equality we used $f_{\pi'} \simeq 0.8 \Lambda'$.

The domain walls dominate the universe around a temperature
\begin{equation}
    T_{\rm dom} \sim \frac{\Lambda^{'3/2}}{\MPl^{1/2}} \sim 10~{\rm keV} \left(\frac{\Lambda'}{10^3~{\rm GeV}}\right)^{3/2},
\end{equation}
in conflict with observations. To avoid this, one must require either that the reheating temperature lie below the mirror QCD scale, or that the domain walls be rendered unstable by explicit breaking of the $Z_3$ symmetry. The latter option is not viable in our setup, since the $Z_3$ symmetry is assumed to be an exact discrete gauge symmetry in order to ensure the quality of the PQ symmetry. 
We are therefore led to require that the maximum temperature of the Universe after inflation and the Hubble scale during inflation be below the mirror QCD scale,%
\footnote{
The maximal temperature of the Universe after inflation is generically much higher than the reheating temperature $T_R$~\cite{Harigaya:2013vwa,Mukaida:2015ria}, so $T_R \ll \Lambda'$ is required.
}
which significantly constrains viable baryogenesis scenarios.

We may also avoid the domain wall problem by embedding the $Z_3$ symmetry into a continuous gauge group, such as $U(1)$. In this case, the symmetry breaking pattern is $U(1)\rightarrow Z_3 \rightarrow $ none, and there will be no absolutely stable domain walls.  Indeed, domain walls can be destructed if a hole bound by a $U(1)$ string is created on them either by the nucleation of string loops on domain walls or by the intersection of domain walls with cosmic strings around~\cite{Kibble:1982ae,Vilenkin:1982ks}. The nucleation rate of strings is exponentially suppressed by the ratio between the string tension $\mu$ and the domain wall tension $\sigma$ as $\sim {\rm exp}(- \mu^{3}/\sigma^2)$~\cite{Kibble:1982dd}, so solving the domain wall problem via nucleation requires that the $U(1)$ symmetry breaking scale be just above the mirror QCD scale. To produce cosmic strings intersecting domain wall requires that $U(1)$ symmetry breaking occur after cosmic inflation. For both cases, the $U(1)$ symmetry breaking scale needs to be much below the cutoff scale of the field theory, and field theoretical UV completion will be required. In the UV completion, anomaly cancellation of the $U(1)$ symmetry will require the introduction of extra particles. The cosmological evolution of those fields should be explored. Also, avoiding the hierarchy problem associated with the $U(1)$ symmetry breaking scale may add extra complication. Therefore, we do not pursue this direction.

\subsection{Stable octet bayon}
\label{sec:baryon}

As we have discussed, the model also predicts $SU(3)_c$-charged stable baryons.
In the limit where the $SU(3)_c$ gauge coupling vanishes, the mirror QCD dynamics reduces to that of a three-flavor QCD. As in the Standard Model, the lightest baryons form an octet representation under the unbroken vector-like $SU(3)_V$ symmetry,
which corresponds to $SU(3)_c$ in this setup.

For later purposes, we briefly explain why the lightest baryons form an octet representation under $SU(3)_c$. Since the baryon is $SU(3)_c'$ singlet, the baryon wavefunction is antisymmetric in the $SU(3)_c'$ indices.
To satisfy the Pauli principle, the wavefunction must be symmetric in the combined $SU(3)_c$ and spin indices. Due to spin--spin interactions, the lightest baryons are expected to have spin $1/2$, which corresponds to a mixed symmetry in the spin indices. Consequently, the $SU(3)_c$ indices must also exhibit mixed symmetry, leading to an octet representation. We denote this lightest baryon by $B_O$.

Because of the domain wall problem, the temperature of the universe should not be above the mirror QCD scale. Therefore, $B_O$ cannot be produced thermally. Still, they can be produced during the thermalization process while the inflaton decays~\cite{Harigaya:2014waa,Harigaya:2016vda,Harigaya:2019tzu,Drees:2022vvn,Mukaida:2022bbo}. We assume such a scenario and discuss the viability of dark matter made from $B_O$.  

Let us assume that the temperature of the universe is above the QCD scale.
After $B_O$ is produced during the thermalization process, its number density is conserved for a while. Around the QCD phase transition, $B_O$ forms bound states with quarks and gluons and scatter with each other with a QCD-scale cross section. ${\cal O}(1)$ fraction of them falls into a deeply bound state of two $B_O$, which is a dark-matter candidate~\cite{DeLuca:2018mzn}, while ${\cal O}(1)$ fraction of them forms $B_O\bar{B}_O$ and annihilates.
By choosing appropriate values of the inflaton mass and the reheating temperature, which determine the abundance of $B_O$, the observed dark-matter abundance can be explained.

A small fraction of $B_O$ remains in bound states with gluons ($B_O g$) or quarks ($B_O q\bar{q}$), which interact strongly with SM hadrons. Constraints on these states were studied in~\cite{DeLuca:2018mzn}, where it was argued that they are allowed provided they do not form electromagnetically charged bound states. This condition may be satisfied if $B_O q\bar{q}$ efficiently decays into $B_O g$, which is an isospin singlet and whose interactions via pion exchange are expected to be suppressed, potentially preventing the formation of electromagnetically charged bound states.

However, Ref.~\cite{DeLuca:2018mzn} overestimates the scattering cross sections of $B_O g$ and $B_O q\bar{q}$ with nuclei~\cite{Digman:2019wdm}, leading to the conclusion that these states cannot reach underground detectors. Using the corrected cross sections, we find instead that these bound states can penetrate the Earth's crust over distances of $\mathcal{O}(1)\,\mathrm{km}$ and reach typical underground detectors if $m_{B_O} \gtrsim 100~\mathrm{TeV}$, in which case the model is excluded. For $m_{B_O} \lesssim 100~\mathrm{TeV}$, the bound states are stopped before reaching detectors, and the model remains viable. $m_{B_O} \lesssim 100~\mathrm{TeV}$ is consistent with the collider lower bound on $\Lambda'$.

Although the existence of stable octet baryons is not immediately excluded, the viability of the scenario relies on an assumption on the mass spectrum. Also, to avoid the domain wall problem requires a low reheating temperature, which limits possible baryogenesis scenarios. In the next section, we present a setup where the domain wall problem is absent and the lightest baryon is color singlet.

\section{Accidental PQ from $U(1)$}
\label{sec:U1}

In this section, we propose another model of accidental PQ symmetry with a mirror SM, where the PQ symmetry arises from a chiral $U(1)$ gauge symmetry. The key difference from the model in Sec.~\ref{sec:Z3} is that the gauge symmetry is continuous rather than discrete; as the result, the model is free from the domain wall problem. Also, we find that SM-charged stable particles are absent in our model.
One of the pseudo-NGBs serves as a dark matter candidate and the chiral $U(1)$ gauge boson, which acquires a mass from the chiral symmetry breaking, provides a vector portal connecting the SM with the dark-matter candidate. Colored pseudo-NGBs can be searched for at the LHC.

\subsection{Setup}

\begin{table}
    \centering
    \begin{tabular}{c|ccc|cc}
         & $SU(3)_c$ & $SU(3)_c'$ & $U(1)_D$ & $U(1)_{\rm PQ}$ & $U(1)_T$ \\ \hline
        $\psi_u$ & ${\bf 3}$ & ${\bf 3}$ & $-a$ & $1/4$ & $1$ \\
        $\psi_d$ & ${\bf 3}$ & ${\bf 3}$ & $a$ & $1/4$ & $-1$ \\
        $\bar{\psi}_u$ & $\overline{{\bf 3}}$ & $\overline{{\bf 3}}$ & $1$ & $1/4$ & $-1$ \\
        $\bar{\psi}_d$ & $\overline{{\bf 3}}$ & $\overline{{\bf 3}}$ & $-1$ & $1/4$ & $1$ \\
    \end{tabular}
    \caption{The quantum numbers of chiral fermions. Here $U(1)_D$ is a gauge symmetry that leads to an accidental $U(1)_{\rm PQ}$ symmetry when $|a| \neq 1$. $U(1)_T$ is another accidental symmetry that leads a stable NGB.}
    \label{tab:charge2}
\end{table}

As in the model discussed in Sec.~\ref{sec:Z3}, we introduce a mirror copy of the SM and the $Z_2$ symmetry exchanging the SM with the SM$'$.
Two pairs of $SU(3)_c\times SU(3)_c'$ fermions shown in Table~\ref{tab:charge2} are introduced and they are charged under a gauge symmetry $U(1)_D$. When $|a|\neq 1$, the mass terms of the fermions are forbidden, and a PQ symmetry arises accidentally. Similar chiral $U(1)$ symmetry is also considered in~\cite{Harigaya:2016rwr,Co:2016akw,Contino:2020god,Ibe:2021gil} where the masses of dark matter and a dark photon are generated by the strong dynamics of a hidden $SU(N)$.

The strong CP problem is resolved in the same manner as in the model described in Sec.~\ref{sec:Z3}. The relevant strong dynamics is that of an $SU(3)$ gauge theory with six flavors, which is expected to undergo chiral symmetry breaking rather than flow to a conformal phase~\cite{Appelquist:2007hu}. The approximate flavor symmetry $SU(6)_L \times SU(6)_R$, where $SU(6)_L$ and $SU(6)_R$ act on $\psi_{u,d}$ and $\bar{\psi}_{u,d}$, respectively, is spontaneously broken to the diagonal subgroup $SU(6)_V$, yielding 35 NGBs. Without loss of generality, we take $a \geq 0$, in which case the following condensate minimizes the vacuum energy:
\begin{equation}
\label{eq:vac}
    \vev{\bar{\psi}_u \psi_u} = \vev{\bar{\psi}_d \psi_d} \sim  \Lambda^{'3}.
\end{equation}
Note that for $a >0$, this is a more attractive channel than $\vev{\bar{\psi}_u \psi_d}= -\vev{\bar{\psi}_d \psi_u} \neq 0$. For $a=0$, there is a moduli space because of the $SU(2)$ symmetry of $\psi_u$ and $\psi_d$, but without loss of generality we can take the vacuum in Eq.~\eqref{eq:vac}.
Baryons made from $\psi_{u,d}$ and $\bar{\psi}_{u,d}$ have masses $\sim$ $ 4 \Lambda'$. See Secs.~\ref{sec:stable} and \ref{sec:collider} for the mass spectrum and phenomenology of the NGBs and baryons. The QCD axion is the $\eta'$-like state of the $SU(3)_{c'}$ dynamics and is as heavy as $4 \Lambda'$.

\begin{figure}
    \centering
    \includegraphics[width=0.7\linewidth]{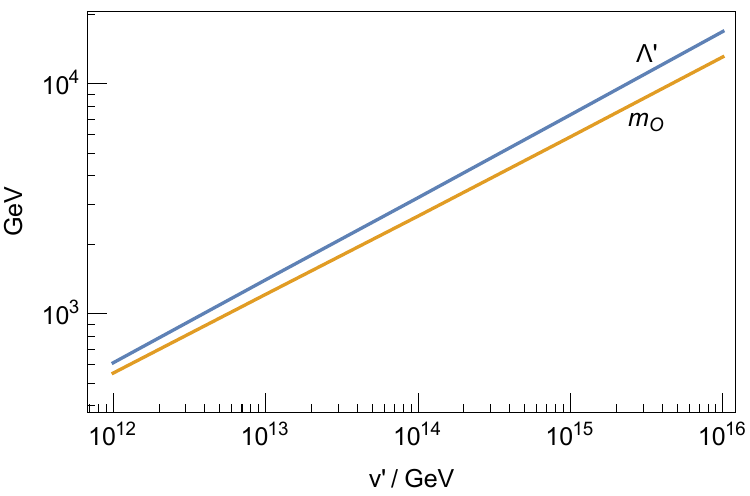}
    \caption{The mirror QCD scale $\Lambda'$ and the mass of octet NGBs as a function of the mirror electroweak scale $v'$ in the model in Sec.~\ref{sec:U1}.}
    \label{fig:vpL}
\end{figure}

In Fig.~\ref{fig:vpL}, we show $\Lambda'$ as a function of $v'$. Here $\Lambda'$ is defined as an energy scale at which the $SU(3)_c'$ gauge coupling computed at the two-loop level diverges. As we will see, there exists color-octet particles with a mass $m_O \simeq 0.8 g_3 \Lambda'$. To evade the LHC bound requires $v'>2 \times 10^{13}$ GeV (see Sec.~\ref{sec:collider}), which is barely consistent with Eq.\,\eqref{eq:dtheta} for $M_H$ around the reduced Planck scale. Future measurements of the neutron and proton electric dipole moments are expected to discover non-zero values.

The $U(1)_{\rm PQ}$ symmetry is explicitly broken by a dimension-6 operator $\psi_u \psi_d \bar{\psi}_u \bar{\psi}_d/M^2$, where $M$ is a mass scale that may be as large as $M_H$. The resultant correction to the $\theta$ term is $O(\Lambda'^2/ M^2)$. To solve the strong CP problem requires that
\begin{equation}
    \Lambda' \lesssim 10^{-5} M = 10^{13}~{\rm GeV} \frac{M}{10^{18}~{\rm GeV}},
\end{equation}
which is satisfied unless $M < 10^{9}$ GeV.

\subsection{Stable relics}
\label{sec:stable}

\subsubsection{Baryons}

In this setup, the lightest baryon is color singlet and is a dark-matter candidate. To derive this,
we decompose $SU(6)_{L,R}$ into $SU(3)_{L,R} \times SU(2)_{L,R}$ such that the diagonal subgroup of $SU(3)_{L,R}$ is $SU(3)_c$ and $(\psi_u,\psi_d)$ and $(\bar{\psi}_u,\bar{\psi}_d)$ are doublets under $SU(2)_{L,R}$, respectively. The unbroken group $SU(6)_V$ is then decomposed as $SU(3)_c\times SU(2)_V$, where $SU(2)_V$ is the diagonal subgroup of $SU(2)_L\times SU(2)_R$.
The statistics of the spin of the lightest baryon is mixed, so $SU(3)_c\times SU(2)_V$ index must also exhibit mixed symmetry to satisfy the Pauli principle. In terms of the original $SU(6)_V$, it is ${\bf 70}$. Decomposing it to $SU(3)_c\times SU(2)_V$, we obtain the representations of the baryons as
\begin{equation}
    ({\bf 1},{\bf 2}) + ({\bf 8},{\bf 2}) + ({\bf 8},{\bf 4}) + ({\bf 10},{\bf 2}).
\end{equation}
In the limit where the $SU(3)_c$ gauge coupling vanishes, these 70 baryons are degenerated. With quantum corrections from the $SU(3)_c$ interaction, $({\bf 1},{\bf 2})\equiv B$ becomes the lightest. Other states decay into the lightest baryon and NGBs.

\subsubsection{Nambu-Goldstone bosons}
With the symmetry breaking pattern $SU(6)_L \times SU(6)_R \rightarrow SU(6)_V$, 35 NGBs arise, transforming in the adjoint representation of the unbroken group $SU(6)_V$. We decompose $SU(6)_V$ into $SU(3)_c\times SU(2)_V$ as described above and the representations of the NGBs are obtained as
\begin{equation}
    (O_{T_0},O_T)({\bf 8},{\bf 3}) + O({\bf 8},{\bf 1}) + (T_0,T)({\bf 1},{\bf 3}).
\end{equation}
Here we decompose $SU(2)_V$ triplets into the analogous of a neutral pion and a charged pion, $O_{T_0}$, $O_{T}$, $T_0$, and $T$. $O_T$ and $T$  are complex fields that are charged under another accidental symmetry $U(1)_T$ shown in Table~\ref{tab:charge2}.
$T_0$ is eaten by the $U(1)_D$ gauge boson and obtains a mass
\begin{equation}
\label{eq:mAD}
    m_{A_D}^2 \simeq 6 e_D^2 (1-a)^2 \frac{m_{\rho'}^2}{16\pi^2},
\end{equation}
where $m_{\rho'}$ is the mass of the $\rho$ meson-like state made from $\psi$ and $e_D$ is the $U(1)_D$ gauge coupling constant.
Other states obtain masses by the $SU(3)_c$ and $U(1)_D$ interactions,
\begin{align}
\label{eq:mNGB}
 m_{O}^2 \simeq & 18 {\rm ln}2g_3^2 \frac{m_{\rho'}^2}{16\pi^2}, ~~
    m_{O_{T}}^2 \simeq  \left(18 {\rm ln}2g_3^2 + 24 a {\rm ln}2 e_D^2  \right)\frac{m_{\rho'}^2}{16\pi^2},   \nonumber \\
    m_{O_{T_0}}^2 \simeq &  18 {\rm ln}2g_3^2 \frac{m_{\rho'}^2}{16\pi^2} ,  ~~
    m_{{T}}^2 \simeq  24 a {\rm ln}2 e_D^2 \frac{m_{\rho'}^2}{16\pi^2}.
\end{align}
Here we modified the results in~\cite{Contino:2020god} that are derived for electroweak charged $\psi$.%
\footnote{
In~\cite{Contino:2020god}, the expression corresponding to $m_{O_{T}}$ misses the factor of $a$.
However, for $g_3=0$ and $a=0$, the symmetry breaking pattern is $SU(6)_L\times SU(3)_{R,u}\times SU(3)_{R,d}\times U(1)_V\times U(1)_D \rightarrow SU(3)_{V,u}\times SU(3)_{V,d} \times U(1)_V\times U(1)_T$, so 
there should be 34 NGBs and one would-be NGB and we conclude the factor of $a$ is necessary.
Here, $SU(3)_{R,u(d)}$ is the $SU(3)$ rotation on $\bar{\psi}_{u(d)}$ and $SU(3)_{V,u(d)}$ is the $SU(3)$ rotation on ${\psi}_{u(d)}$ and $\bar{\psi}_{u(d)}$.
}
In Fig.~\ref{fig:vpL}, we show the mass of $O$ as a function of $v'$. Here we assume $\Lambda'/m_{\rho'} = \Lambda/m_\rho =(260~{\rm MeV})/(770~{\rm MeV})$. 
Note that the estimation of the masses assumes so-called the vector dominance~\cite{Sakurai:1960ju,Das:1967it}, which is accurate in large-$N_c$ limit (e.g.,~\cite{Masjuan:2012sk}.) In the current setup, the number of flavor is six and it is not clear whether
the large-$N_c$ limit is applicable. This introduces uncertainty when we translate the bound on the NGB masses into that on $v'$.
The ratios of the NGB masses are free from such uncertainty.  

Octet NGBs, $O$, $O_T$, and $O_{T_0}$, acquire masses from the $SU(3)_c$ gauge interaction regardless of the value of $a$. Therefore, they are unstable and expect to decay. Their decay channels are discussed in Appendix~\ref{app:discrete symmetry}, where we find that
$O$ decays into a pair of gluons while
$O_T$ and $O_{T_0}$ decays into gluons, $T$, and $A_D$.

$A_D$ can decay into $T$ and $\bar{T}$ if kinematically allowed, which is possible for sufficiently small $a$. If not, by introducing kinetic mixing between $U(1)_D$ and $U(1)_Y$, it can decay into a pair of SM particles.%
\footnote{$A_D$ may also decay into $O_{T_0} g$ via the Wess-Zumino-Witten term\,\cite{Witten:1983tw} if kinematically allowed. This requires $e_D > g_3$ and leads to the Landau pole of $e_D$ much below $v'$.}
The kinetic mixing is not generated by loop corrections. This is because the $U(1)_D$ interaction preserves a charge-conjugation symmetry, $C_D$, under which the fields transform as $A_D \rightarrow -A_D $, $\psi_u \leftrightarrow \psi_d$, and $\bar{\psi}_u \leftrightarrow \bar{\psi}_d$ as discussed in Appendix~\ref{app:discrete symmetry}. Therefore, the kinetic mixing needs to be introduced at tree-level.

For $a = 0$, $T$ remains massless.
This can be understood from the symmetry structure of the model. The exact symmetry of the model for $a=0$ is $SU(3)_c \times SU(2)_L \times U(1)_D \times U(1)_V$, where $U(1)_V$ is the vector-like symmetry under which all the fermions have the same charge. The symmertry is spontaneously broken to $SU(3)_c \times U(1)_\text{diag} \times U(1)_V$, where $U(1)_\text{diag}$ is the diagonal subgroup of $SU(2)_L$ and $U(1)_D$. Therefore, there are three NGBs, one of which is eaten by the $U(1)_D$ gauge boson and the other two form a massless complex scalar $T$.
$T$ contributes to dark radiation of the universe. They decouple from the thermal bath before the electroweak phase transition, so $\Delta N_{\rm eff}\simeq 0.054$. The global symmetry $SU(2)_L$ may be explicitly broken by higher-dimensional operators, such as $\psi_u^2 \bar{\psi}_u \bar{\psi}_d/M^2$, which gives a mass of
\begin{equation}
    m_T \sim \frac{\Lambda^{'2}}{M} =  0.01~{\rm eV} \left(\frac{\Lambda'}{{\rm TeV}} \right)^2 \frac{10^{17}~{\rm GeV}}{M}
\end{equation}
to $T$.
As we will see in Sec.~\ref{sec:DW}, destroying the domain wall before the BBN requires $M < 10^{17}$ GeV $\left(\Lambda'/{\rm TeV}\right)^{3/2}$. 
The resultant mass of $T$ is sufficiently small to satisfy the Planck constraint~\cite{Planck:2018vyg}.

For $a\neq 0$, $T$ becomes massive. Because of the accidental $U(1)_T$ symmetry, $T$ is stable and may be dark matter~\cite{Harigaya:2016rwr,Co:2016akw,Contino:2020god}.
Indeed, for generic $a$, there are no $U(1)_T$-violating interactions with dimension below $7$. For some values of $a$, dimension-7 operators are allowed. For example, for $a=1/3$, $\psi_u^2 \psi_d^\dag \slashed{D}\bar{\psi}_u $ is allowed by the gauge symmetry. However, the resultant lifetime of $T$ is much longer than the lower bound from indirect detection. 
The relic abundance of $T$ is determined by the freeze out of the annihilation through the $U(1)_D$ interaction. See Sec.~\ref{sec:DM} for dark-matter phenomenology. 

\subsubsection{Mirror quarks and electrons}

The mirror electrons and the mirror hadrons containing $u'$ or $d'$ may be stable.
In fact, if all the gauge groups of the SM is mirrored and the gauge symmetry is $SU(3)_c\times SU(3)_c'
\times SU(2)_w\times SU(2)_w'\times U(1)_Y \times U(1)_Y'$, $e'$ is absolutely stable because of the mirror electromagnetic charge conservation. In the viable parameter region where $v'>10^{13}$ GeV, $m_e' > 10^7$ GeV and the freeze-out abundance of $e'$ is too large. By assuming reheating temperature much below $0.1 m_{e'}$, we may avoid this problem.

Alternatively, while keeping a high reheating temperature, we may consider the gauge group $SU(3)_c\times SU(3)_c'
\times SU(2)_w\times SU(2)_w'\times U(1)_X$. The gauge quantum numbers of the SM fermions, Higgs, and their mirror copies are shown in Table~\ref{tab:charge3}. $SU(2)_w'\times U(1)_X$ is broken down into $U(1)_Y$ by non-zero $\vev{H'}$.
The mirror electron can decay via a mass term $\bar{e} \bar{e}'$. If the mass term is large, we may integrate out $\bar{e}$ and $\bar{e}'$ and the $U(1)_Y$-charged component of $\ell'$ is the SM right-handed charged leptons.  

If $u'$ or $d'$ remain stable, their relic abundance determined by the freeze-out mechanism is too large, so they should be destabilized.
For $a\neq0$, $u'$ and $d'$ can decay via the following interactions
\begin{equation}
\label{eq:decay1}
    \frac{1}{{M'}^2} \left( \bar{u}' \bar{u} \bar{\psi}_u \bar{\psi}_d  + \bar{d}' \bar{d} \bar{\psi}_u \bar{\psi}_d  \right).  
\end{equation}
For $a=0$, the following interactions are also allowed by gauge symmetry and let $u'$ and $d'$ decay,
\begin{equation}
\label{eq:decay2}
    \frac{1}{{M'}^2} \left( \bar{u}' \psi \bar{d} \bar{e}'  +  \bar{d}' \psi \bar{u} \bar{e} + q' \ell' \bar{d}^\dag \psi^\dag  \right) . 
\end{equation}
The temperature around which $u'$ and $d'$ decay is
\begin{equation}
\label{eq:Tdec}
 T_{\rm dec} \simeq 1~{\rm TeV} \left(\frac{10^{12}~{\rm GeV}}{M'}\right)^2 \left(\frac{v'}{10^{13}~{\rm GeV}}\right)^{5/2}.
\end{equation}
If the decay occurs before the mirror QCD phase transition, $\psi$ produced by the decay of $u'$ and $d'$ are thermalized and the abundance of $B$ and $T$ are determined by the standard freeze-out mechanism discussed in Sec.~\ref{sec:DM}.%
\footnote{Beta decay of $d'$ by $W'$ exchange occurs after the mirror QCD phase transition, so the second term in Eq.~\eqref{eq:decay1} or the second or third term in Eq.~\eqref{eq:decay2} needs to be introduced to let $d'$ decay early enough.}
This requires extra particles below the scale $v'$ to UV-complete the interactions in Eqs.~\eqref{eq:decay1} or \eqref{eq:decay2}.
If the decay occurs after the freeze out of the annihilation of $B$ and $T$, the abundance of them can be larger than that determined by the freeze-out mechanism. As we will discuss in Sec.~\ref{sec:baryon DM}, this enables $B$ to be the dominant component of dark matter.

\begin{table}
    \centering
    \begin{tabular}{c|ccccc}
         & $SU(3)_c$  & $SU(3)_c'$  & $SU(2)_w$  & $SU(2)_w'$  & $U(1)_X$ \\ \hline
       $q$ & ${\bf 3}$  & ${\bf 1}$  & ${\bf 2}$  & ${\bf 1}$  & $1/6$ \\
       $q'$ & ${\bf 1}$  & ${\bf 3}$  & ${\bf 1}$  & ${\bf 2}$  & $-1/6$ \\
       $\bar{u}$  & ${\bf \bar{3}}$  & ${\bf 1}$  & ${\bf 1}$  & ${\bf 1}$  & $-2/3$ \\
    $\bar{u}'$  & ${\bf 1}$ & ${\bf \bar{3}}$     & ${\bf 1}$  & ${\bf 1}$  & $2/3$ \\
      $\bar{d}$  & ${\bf \bar{3}}$  & ${\bf 1}$  & ${\bf 1}$  & ${\bf 1}$  & $1/3$ \\
    $\bar{d}'$  & ${\bf 1}$ & ${\bf \bar{3}}$     & ${\bf 1}$  & ${\bf 1}$  & $-1/3$ \\
    $\ell$ & ${\bf 1}$ & ${\bf 1}$ & ${\bf 2}$ & ${\bf 1}$ & $-1/2$ \\
    $\ell'$ & ${\bf 1}$ & ${\bf 1}$ & ${\bf 1}$ & ${\bf 2}$ & $1/2$ \\
    $\bar{e}$ & ${\bf 1}$ & ${\bf 1}$ & ${\bf 1}$ & ${\bf 1}$ & $1$ \\
    $\bar{e}'$ & ${\bf 1}$ & ${\bf 1}$ & ${\bf 1}$ & ${\bf 1}$ & $-1$ \\
    $H$ & ${\bf 1}$ & ${\bf 1}$ & ${\bf 2}$ & ${\bf 1}$ & $1/2$ \\
    $H'$ & ${\bf 1}$ & ${\bf 1}$ & ${\bf 1}$ & ${\bf 2}$ & $-1/2$ \\
    \end{tabular}
    \caption{Gauge charges of the SM fermions, Higgs, and their mirror partners.}
    \label{tab:charge3}
\end{table}

\subsection{Topological defects}
Similar to the model in Sec.~\ref{sec:Z3}, our model admits several topological defects. Let us discuss the symmetry structure of the model to systematically understand the topological defects.

First, let us consider the limit where $g_3=0$ and $e_D=0$, i.e., turn off the $SU(3)_c$ and $U(1)_D$ gauge interactions. In this limit, the theory has a global symmetry $G_\text{approx} \equiv SU(6)_L\times SU(6)_R\times U(1)_V$, with the $U(1)_\text{PQ}$ explicitly broken by the $SU(3)_{c'}$ anomaly. 
As we have discussed, after the chiral symmetry breaking, the symmetry is broken down to $H_\text{approx} \equiv SU(6)_V\times U(1)_V$. The coset space is diffeomorphic to $SU(6)$ and no stable topological defects exist. 

Next, let us turn on the gauge couplings. First, we turn on $g_3$. 
To see the symmetry of the theory, it is again useful to decompose $SU(6)_{L,R}$ into $SU(3)_{L,R} \times SU(2)_{L,R}$.  
If we gauge $SU(3)_c$, the $SU(3)$s in $SU(6)_L$ and $SU(6)_R$ are explicitly broken, and only the diagonal subgroup $SU(3)_c$ and the discrete subgroup $Z_3$, the axial rotation of the central element of each $SU(3)$, remain as symmetries. Therefore,
the symmetry of the system is $G_\text{approx}' \equiv SU(3)_c\times Z_3 \times SU(2)_L\times SU(2)_R \times U(1)_V$, which is broken down to $H_\text{approx}' \equiv SU(3)_c \times SU(2)_V \times U(1)_V$ by the chiral symmetry breaking.

Finally, we turn on $e_D$. The $U(1)_D$ gauge symmetry is a subgroup of $SU(2)_L$ and $SU(2)_R$. It depends on the value of $a$ which subgroup is gauged. For $a=0$, the $U(1)_D$ is the Cartan subgroup of the $SU(2)_R$. In this case, $SU(2)_L$ is intact and remains as a symmetry of the system, while $SU(2)_R$ is explicitly broken to $U(1)_D$. 
The symmetry of the system is $G_\text{exact} \equiv SU(3)_c\times Z_3 \times SU(2)_L\times U(1)_D \times U(1)_V$ and  is broken down to $H_\text{exact} \equiv SU(3)_c \times U(1)_\text{diag} \times U(1)_V$.  
For $a\ne 0$, $U(1)_D$ is a linear combination of the Cartan subgroups of $SU(2)_L$ and $SU(2)_R$. In this case, both $SU(2)_L$ and $SU(2)_R$ are explicitly broken to their Cartan subgroups. The symmetry of the system is $G_\text{exact} \equiv SU(3)_c\times Z_3 \times U(1)_D \times U(1)_T \times U(1)_V$, which is broken down to $H_\text{exact} \equiv SU(3)_c \times U(1)_\text{diag} \times U(1)_V$.

With this symmetry breaking pattern, the following topological defects are produced. First, in either case of $a=0$ and $a\neq 0$, $Z_3$ subgroup of $G_\text{exact}$ is spontaneously broken, so $\pi_0(G_\text{exact}/H_\text{exact}) = Z_3$ and domain walls can be produced. In addition, for $a\ne 0$, $\pi_1(G_\text{exact}/H_\text{exact}) = Z$ and cosmic strings  can be produced. For $a=0$, the situation is slightly subtle. Although $\pi_1(G_\text{exact}/H_\text{exact}) = 0$, if we focus on the gauged subgroup of the symmetry, $G_\text{exact} \supset G_\text{gauge} \equiv SU(3)_c\times U(1)_D$ and $H_\text{exact} \supset H_\text{gauge} \equiv SU(3)_c$, so we have $\pi_1(G_\text{gauge}/H_\text{gauge}) = Z$. In such cases, whether string solutions exist or not depends on the size of parameters in the theory\,\cite{Preskill:1992bf,Vachaspati:1991dz}. These strings are called semi-local strings if exist. 
Below we discuss the domain wall and string solutions in detail.
In addition, throughout this paper, we assume that any global symmetries are explicitly broken by higher-dimensional operators. 
We find that, after including higher-dimensional operators, the domain wall becomes unstable while a string solution exists even for $a=0$.

\subsubsection{Metastable domain walls and gravitational waves}
\label{sec:DW}
Let us first explicitly show the existence of a stable domain wall solution when higher-dimensional interactions are not introduced.
One may naively extend the discussion in Sec.~\ref{sec:Z3} based on $Z_3$ into $Z_{3,u}\times Z_{3,d}$ and expect the following vacuum structure,
\begin{equation}
    \vev{\psi_u\bar{\psi_u}}\sim \Lambda^{'3}{\rm exp}\left( i \frac{2\pi}{3}k_u\right),~~\vev{\psi_d\bar{\psi_d}}\sim \Lambda^{'3}{\rm exp}\left( i \frac{2\pi}{3}k_d\right),~~k_u,k_d = 0,1,2.
\end{equation}
It looks like that there are nine disconnected vacua, which is inconsistent with the discussion at the beginning of this section indicating that the domain wall number is three. This is because a part of $Z_{3,u}\times Z_{3,d}$ is a subgroup of $U(1)_D$ and some of the nine points are continuously connected.
Let us consider the following configuration that connects $(k_u,k_d)=(0,0)$ with $(1,-1)$,
\begin{equation}
    \vev{\psi_u\bar{\psi_u}}\sim \Lambda^{'3}{\rm exp}\left( i \frac{2\pi}{3}f(z)\right),~~\vev{\psi_d\bar{\psi_d}}\sim \Lambda^{'3}{\rm exp}\left( -i \frac{2\pi}{3}f(z)\right),~~ f(-\infty)=0,~f(\infty)=1.
\end{equation}
Because of the $U(1)_D$ symmetry under which $\psi_u\bar{\psi}_u$ and $\psi_d\bar{\psi}_d$ have opposite charges,%
\footnote{
This is in contrast to a solution to the $Z_3$ domain wall problem discussed in Sec.~\ref{sec:Z3_DW}, where the $U(1)$ symmetry breaking field has a $U(1)$ charge of $3$ while $\psi \bar{\psi}$ has a $U(1)$ charge of unity, for which classically stable domain walls associated with $Z_3 \subset U(1)$ exist. In the $U(1)_D$ model, because the absolute values of the $U(1)_D$ charges of  $\vev{\psi_u\bar{\psi_u}}$ and  $\vev{\psi_d\bar{\psi_d}}$ are equal, none of them breaks $U(1)_D$ down to a discrete subgroup. Therefore, classically stable domain walls associated with a discrete subgroup of $U(1)_D$ symmetry does not exist. 
}
there is no potential energy along this path.
Also, the gradient energy can be eliminated by taking $A_z\propto f'(z)$. Generically, $(k_u,k_d)= (m,n)$ and $(m+1,n-1)$ are continuously connected by $U(1)_D$ without an energy barrier, and only the three vacua with $k_u =  k_d$ are disconnected.
Equivalently,
to remove the effect of the $U(1)_D$ symmetry, we may take an order parameter that is neutral under $U(1)_D$,
which is $\psi_u \psi_d \bar{\psi}_u \bar{\psi}_d$. This order parameter takes on three values
\begin{equation}
    \vev{\psi_u \psi_d \bar{\psi}_u \bar{\psi}_d} \sim \Lambda^{'6} {\rm exp} \left( i \frac{2\pi}{3}k\right),~~k=0,1,2 
\end{equation}
at the vacua.

These three vacua are connected by the axial rotation of the central element of $SU(3)_L$ and $SU(3)_R$, as discussed in Sec.~\ref{sec:Z3}. The potential barrier between these three vacua is generated by the $SU(3)_c$ gauge interaction, and therefore, after turning on $SU(3)_c$ gauge coupling, the theory indeed has three distinctive disconnected vacua and the domain wall number is three, which is consistent with the discussion at the beginning of this section. 
The domain walls are made from a component of $O^8$, and thus have a tension $\simeq 1.7 g_3 \Lambda^{'3}$, similar to the model in Sec.~\ref{sec:Z3}.
Inside the domain walls, $SU(3)_c$ is again broken down to $SU(2)\times U(1)$.

Up to this point, we have assumed that the symmetry group, $G_\text{exact}$, is exact. In reality, however, it is expected that there are higher-dimensional operators that explicitly break the global symmetry part of $G_\text{exact}$.
Such breaking lifts the degeneracy of the three vacua and makes the domain walls unstable. 
More specifically, the most relevant operator allowed by the gauge symmetry is $\psi_u \psi_d \bar{\psi}_u \bar{\psi}_d/M^2$, and the generated difference of the energy density of the three local minima is $\Lambda^{'6}/M^2$.%
\footnote{$\Lambda'$ is the scale at which the mirror QCD coupling diverges, which corresponds to $\Lambda\sim 260$ MeV in the QCD. The quark bilinear condensation scale in the three-flavor QCD is about $\Lambda^3$.
In the current setup, the number of flavor is six and there might be some difference between the two scales.}
Domain walls decay when this is as large as $\sigma H$. The Hubble scale and the temperature at the time of decay is
\begin{equation}
    H_{\rm dec} \simeq \frac{\Lambda^{'3}}{1.7 g_3M^2},~~
    T_{\rm dec} \simeq 1~{\rm MeV} \left(\frac{\Lambda'}{2~{\rm TeV}}\right)^{3/2} \frac{10^{17}~{\rm GeV}}{M}.
\end{equation}
The energy density of domain walls is $2A \sigma H$, where $A\simeq 0.8$~\cite{Hiramatsu:2013qaa}.
The energy density divided by the entropy density when domain walls decay is
\begin{equation}
\label{eq:rhosDW}
    \frac{\rho_{\rm DW}}{s} \simeq 1 \times 10^{-6}~{\rm GeV} \left(\frac{\Lambda'}{2~{\rm TeV}}\right)^{3/2} \frac{M}{10^{17}~{\rm GeV}}.  
\end{equation}
We expect that the domain walls dominantly decay into $O$, as it mainly constitutes the domain walls.
$O$ then decays into gluons. To avoid disturbing the BBN, the domain walls should decay at $t< 0.1$ sec~\cite{Kawasaki:2017bqm}, which requires
\begin{align}
    M <  4 \times 10^{16}~{\rm GeV}  \left(\frac{\Lambda'}{2~{\rm TeV}}\right)^{3/2}.
\end{align}
Even for $M$ close to the gravitational scale, domain walls can decay before the BBN.

The domain walls produce gravitational waves.
The energy density of gravitational waves while domain walls are present is $\epsilon A^2 \sigma^2 / (8\pi \MPl^2)$ with $\epsilon \simeq 0.7$~\cite{Hiramatsu:2013qaa}. Converting this into the current energy fraction, we obtain
\begin{align}
    \Omega_{\rm GW}h^2& \simeq  2\times 10^{-13} \frac{\epsilon}{0.7} \left( \frac{A}{0.8}\right)^2 \left(\frac{M}{10^{17}~{\rm GeV}}\right)^4
\end{align}
with a peak frequency
\begin{align}
    f_{\rm peak} &\simeq  0.02~{\rm nHz} \left(\frac{\Lambda'}{2 ~{\rm TeV}}\right)^{3/2} \frac{10^{17}~{\rm GeV}}{M}.
\end{align}
The peak frequency is too low to detect even with pulsar timing arrays, but high-frequency tail, which has $\Omega_{\rm GW}(f) \propto f^{-1}$~\cite{Hiramatsu:2013qaa,Kitajima:2023cek}, may be detectable.

\subsubsection{Semi-local cosmic strings}

The theory also contains string solutions. For $a\neq 0$, the relevant part of the continuous symmetry breaking pattern is $G_{\rm exact}=U(1)_D\times U(1)_{\rm diag} $ down to $H_{\rm exact}= U(1)_{\rm diag} $, so cosmic strings are produced by the mirror QCD phase transition. The structure of the string, however, is different from usual localized strings~\cite{Preskill:1992bf,Vachaspati:1991dz}. In the limit where $e_D \rightarrow 0$, the symmetry breaking is $G_{\rm approx}=SU(2)_L\times SU(2)_R$  down to $H_{\rm approx} = SU(2)_V$, and there are no stable string solutions. This is because strings are unwound by the excitation of the $G_{\rm approx}/H_{\rm approx}$ directions. With $e_D \neq 0$, $T$ obtains masses by quantum corrections, and cosmic strings are stabilized, as the excitation of $T$ with sufficiently large radius is energetically disfavored by the potential energy of $T$.
The string tension is ${\cal O}(\Lambda^{'2})$.

For $a=0$, the continuous symmetry breaking pattern is $G_{\rm exact}=SU(2)_L\times U(1)_D$ into $H_{\rm exact}=U(1)_{\rm diag}$, so there do not seem to be string solutions. In fact, the would-be singular point of the cosmic string, i.e., the center of the core where the quark condensate vanishes, can be resolved by the $SU(2)_L$ directions. Then, the winding configuration of the condensation can be spread out to an arbitrary large region, which is a similar structure to the case of the cosmic texture\,\cite{Turok:1989ai}. In addition, the would-be cosmic string also contains the $U(1)_D$ magnetic flux, which prefers to be spread out to an infinite region.
    It is then a dynamical question which configuration, the localized string or the infinitely spread-out configuration, is energetically favored\,\cite{Hindmarsh:1991jq}.
    The stable/unstable boundary is given by the Bogomol'nyi bound, which is determined by the ratio of the scalar self-coupling and the gauge coupling in the perturbative setup with scalar Higgs fields\,\cite{Vachaspati:1991dz}.
    In our case, the condensate comes from strong dynamics but the $U(1)_D$ gauge coupling is perturbative, so we expect that the infinitely spread-out configuration is energetically favored and there is no string solution.

However, $SU(2)_L$ is explicitly broken by a higher dimensional operator $\psi \psi \bar{\psi}_u\bar{\psi}_d$, and the actual symmetry breaking is $G_{\rm exact}=U(1)_D $ down to nothing, and there should be a string solution.
In fact, as the magnetic flux and the excitation of the $SU(2)_L$ direction spread out, once the potential energy of the $SU(2)_L$ direction becomes non-negligible, the configuration stops spreading and is stabilized.  The tension of the string is dominantly given by the gradient energy of the $SU(2)_L$ direction and ${\cal O}(\Lambda^{'2})$.

For both cases, the string tension is ${\cal O}(\Lambda^{'2})$. In the viable parameter space where $\Lambda' = {\cal O}(1)$ TeV, the gravitational waves emitted from cosmic strings are too weak to be detected.
On the other hand, apart from our current model, these types of spread cosmic strings might be produced in a wide class of theories with chiral $U(1)$ gauge symmetry such as dynamical spontaneous $B-L$ breaking models. For such theories, there can be exotic observational consequences; the string tension and the efficiency of loop production can depend on time, which affects the spectrum of the gravitational waves emitted from the string network.%
\footnote{
For $a=0$, there exist nearly massless NGBs that couple to strings. However, the emission of NGBs from string loops is suppressed, and the string loops lose their energy dominantly by the emission of gravitational waves~\cite{Fukuda:2026zoa}.
}

\subsection{Collider signals}
\label{sec:collider}

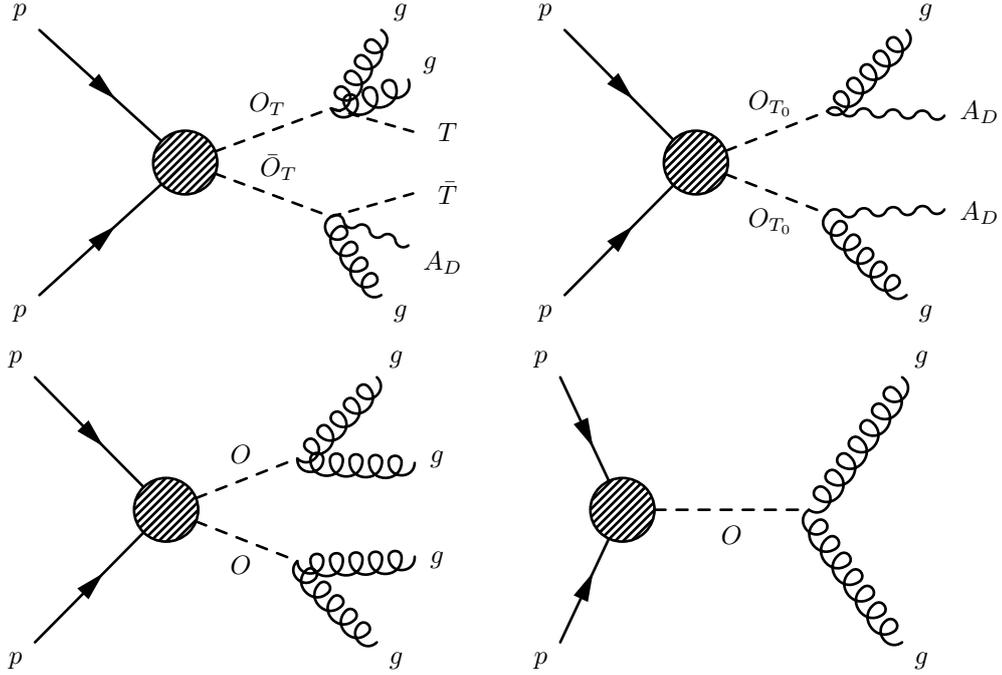
\begin{figure}
    \centering
    \begin{fmffile}{feyngraph1}
  \begin{fmfgraph*}(160,100)
    \fmfleft{i1,i2}
    \fmflabel{$p$}{i1}
    \fmflabel{$p$}{i2}
    \fmfright{o1,o2,o3,o4,o5,o6}
    \fmflabel{$g$}{o1}
    \fmflabel{$A_D$}{o2}
    \fmflabel{$\bar{T}$}{o3}
    \fmflabel{$T$}{o4}
    \fmflabel{$g$}{o5}
    \fmflabel{$g$}{o6}
    \fmf{fermion,tension=1.5}{i1,v1}
    \fmf{fermion,tension=1.5}{i2,v1}
    \fmf{dashes,tension=1.5,label=$\bar{O}_{T}$}{v1,v2}
    \fmf{dashes}{v2,o3}
    \fmf{gluon}{v2,o1}
    \fmf{photon}{v2,o2}
    \fmf{dashes,tension=1.5,lab.side=left,label=$O_{T}$}{v1,v3}
    \fmf{dashes}{v3,o4}
        \fmf{gluon}{v3,o5}
    \fmf{gluon}{v3,o6}
    \fmfblob{0.15w}{v1}
  \end{fmfgraph*}
  \end{fmffile}
  \hspace{2em}
  \begin{fmffile}{feyngraph2}
   \begin{fmfgraph*}(160,100)
    \fmfleft{i1,i2}
    \fmflabel{$p$}{i1}
    \fmflabel{$p$}{i2}
    \fmfright{o1,o2,o5,o6}
    \fmflabel{$g$}{o1}
    \fmflabel{$A_D$}{o2}
    \fmflabel{$A_D$}{o5}
    \fmflabel{$g$}{o6}
    \fmf{fermion,tension=1.5}{i1,v1}
    \fmf{fermion,tension=1.5}{i2,v1}
    \fmf{dashes,tension=1.5,lab.side=right,label=$O_{T_0}$}{v1,v2}
    \fmf{gluon}{v2,o1}
    \fmf{photon}{v2,o2}
\fmf{dashes,tension=1.5,lab.side=left,label=$O_{T_0}$}{v1,v3}
    \fmf{photon}{v3,o5}
    \fmf{gluon}{v3,o6}
    \fmfblob{0.15w}{v1}
  \end{fmfgraph*}
  \end{fmffile}
  \\[3em]
  \begin{fmffile}{feyngraph3}
  \begin{fmfgraph*}(160,100)
    \fmfleft{i1,i2}
    \fmflabel{$p$}{i1}
    \fmflabel{$p$}{i2}
    \fmfright{o1,o2,o3,o4}
    \fmflabel{$g$}{o1}
    \fmflabel{$g$}{o2}
    \fmflabel{$g$}{o3}
    \fmflabel{$g$}{o4}
    \fmf{fermion,tension=1.5}{i1,v1}
    \fmf{fermion,tension=1.5}{i2,v1}
    \fmf{dashes,tension=1.5,label=$O$}{v1,v2}
    \fmf{gluon}{v2,o2}
    \fmf{gluon}{v2,o1}
    \fmf{dashes,tension=1.5,lab.side=left,label=$O$}{v1,v3}
    \fmf{gluon}{v3,o3}
    \fmf{gluon}{v3,o4}
    \fmfblob{0.15w}{v1}
  \end{fmfgraph*}
\end{fmffile}
\hspace{2em}
 \begin{fmffile}{feyngraph4}
  \begin{fmfgraph*}(160,100)
    \fmfleft{i1,i2}
    \fmflabel{$p$}{i1}
    \fmflabel{$p$}{i2}
    \fmfright{o1,o2}
    \fmflabel{$g$}{o1}
    \fmflabel{$g$}{o2}
    \fmf{fermion,tension=1.5}{i1,v1}
    \fmf{fermion,tension=1.5}{i2,v1}
    \fmf{dashes,tension=1,label=$O$}{v1,v2}
    \fmf{gluon}{v2,o2}
    \fmf{gluon}{v2,o1}
    \fmfblob{0.15w}{v1}
  \end{fmfgraph*}
\end{fmffile}
   \caption{Collider signatures of the octet pseudo Nambu-Goldstone bosons $O_{T}$, $O_{T_0}$, and $O$. The state $T$ is invisible and contributes to missing energy. The dark photon $A_D$ decays into a pair of $T$ if kinematically allowed, and thus also contributes to missing energy. Otherwise, it decays into a pair of quarks, leptons, $W^+W^-$, or $Zh$ via kinetic mixing with the hypercharge gauge boson. Depending on the magnitude of this mixing, $A_D$ can produce prompt jets or leptons, displaced vertices, or missing energy.}
    \label{fig:collider}
\end{figure}

The model predicts color-octet NGBs $O_T$, $O_{T_0}$, and $O$, which can be produced at collider experiments. In this section we discuss the LHC phenomenology of the octets, which is summarized in Fig.~\ref{fig:collider}.

The octet NGBs are pair-produced by proton-proton collision. This occurs through the $SU(3)_c$ gauge interaction as well as through the decay of $\rho$-meson like states that are singly produced by the annihilation of a quark and an anti-quark. Since the $\rho$-meson like states have masses $\sim 3 \Lambda' > 2 m_O$, the total cross section is dominated by the former channel, although the latter can create a bump in the invariant-mass distribution of pairs of octet NGBs~\cite{Kilic:2009mi}.

$O_{T}$ dominantly decays into $T gg$ or $T g A_D$.
Since $T$ is stable, the signal is jets plus missing energy. The production cross section of $O_{T}$~\cite{ATLAS:2017jnp} happens to be similar to that of eight squarks~\cite{Beenakker:2024jwh}, and the constraint on $O_{T}$ can be derived by reinterpreting squark searches that put a constraint on the mass of eight degenerate squarks~\cite{ATLAS:2020syg,ATLAS:2021kxv}.  
The constraint depends on the mass spectrum. For $e_D < g_3$, the mass splitting between $O_T$ and $T$ is large and the constraint can be as strong as $m_{O_T} > 1.8$ TeV.  If $e_D > g_3$, the mass splitting is small and the constraint can be relaxed down to 800 GeV. 
However, $e_D>g_3$ leads to a Landau pole of the $U(1)_D$ gauge coupling even at a scale below $v'$.
To avoid the Landau pole below the Plank scale requires $e_D < 0.5$, for which $m_{T} / m_O < 0.5$ and the current constraint is $m_{O_T} > 1.5$ TeV.

$O_{T_0}$ is pair-produced and decays into $A_D g$. If $A_D$ dominantly decays into $T$, the signal is two jets plus missing and the squark search is applicable. If $A_D$ dominantly decays into SM particles via kinetic mixing, missing energy may be absent. As we will see in Sec.~\ref{sec:DM}, when $T$ is dark matter, $m_{T} >m_{A_D}$ in the viable parameter space and $A_D$ cannot decay into $T$.

The signal in this case depends on the lifetime of $A_D$, which is
\begin{equation}
\label{eq:lifetime}
    c \tau = \left(\frac{41\epsilon^2e^2}{96\pi c_w^4} m_{A_D}\right)^{-1} \simeq 1~{\rm cm} \left(\frac{10^{-7}}{\epsilon}\right)^2 \frac{100~{\rm GeV}}{m_{A_D}},
\end{equation}
where $\epsilon$ is the kinetic mixing between $A_D$ and the photon. Here we assume that $m_{A_D}$ is above the electroweak scale, although the lifetime remains of the same order as Eq.~\eqref{eq:lifetime} even for smaller $m_{A_D}$ unless $m_{A_D} < 2 m_e$. 
For $c \tau \lesssim 1$ mm, $A_D$ decays promptly, and multi-jets and/or multi-leptons searches can put constraints on $m_{O_{T_0}}$. Reinterpreting the gluino search with R-parity violation~\cite{ATLAS:2024kqk}, we deduce $m_{O_{T_0}} > 800$ GeV, which is weaker than the constraint from $O_{T}$. However, the search does not take the invariant mass of leptons. In our model, since di-lepton invariant masses are peaked at $m_{A_D}$, a dedicated search using di-lepton invariant masses might put a stronger constraint.
For $1$ mm $\lesssim c \tau \lesssim 1$ m, displaced-vertex search can put a stronger constraint than the jets plus missing energy search. Note that there will be two displaced vertices per event. If the signal is background-free and the efficiency is nearly $100$\%, $m_O < 2300$ GeV and 3000 GeV  can be probed with the integrated luminosity of $100$ fb$^{-1}$ and $3$ ab$^{-1}$ with $\sqrt{s}=14$ TeV, respectively. Here we require that more than five events be expected.
For $c \tau \gtrsim 1$ m, jet plus missing search will be applicable. 

$O$ is also pair-produced and decays into a pair of jets via
\begin{equation}
\label{eq:WZW}
    \frac{3g_3^2}{32 \pi^2} d^{abc}\frac{O^a}{f_\pi'} \epsilon^{\mu \nu \rho \sigma} G^b_{\mu \nu} G^c_{\rho \sigma},
\end{equation}
where $f_\pi'$ is the decay constant of the pseudo NGBs.
The LHC constraint based on this process is $m_O>800$ GeV~\cite{ATLAS:2017jnp}.

\begin{figure}
    \centering
    \includegraphics[width=0.7\linewidth]{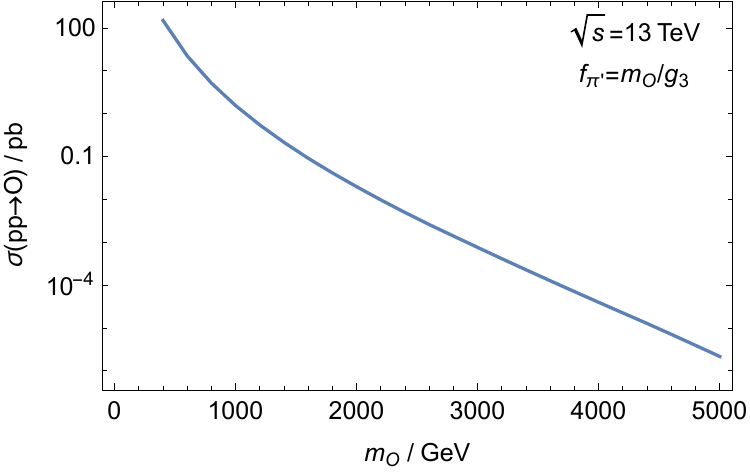}
    \caption{The single production cross section of $O$ in proton-proton collisions.}
    \label{fig:Xsec}
\end{figure}

$O$ can be also singly produced by the interaction in Eq.~\eqref{eq:WZW}. The parton level cross section is
\begin{equation}
    \hat{\sigma}(gg \rightarrow O) = \frac{\pi}{m_O^2} \frac{15 \alpha_3^2 m_O^4}{64 \pi^2 f_{\pi'}^2}\delta (\hat{s}-m_O^2).
\end{equation}
The hadron level cross section is
\begin{equation}
   \sigma (pp \rightarrow O) = K\times \frac{15 \alpha_3^2}{64 \pi} \frac{m_O^2}{f_{\pi'}^2 s} \int \frac{dx}{x} f_g(x) f_g(\frac{m_O^2}{x s}),
\end{equation}
where $K$ is the $K$-factor, which takes into account the difference between the leading and higher-order computations, and $f_g$ is the parton distribution function of the gluon. In Fig.~\ref{fig:Xsec}, we show the production cross section of $O$ at $\sqrt{s}= 13$ TeV with the NNPDF3.0~\cite{NNPDF:2014otw}. To determine $f_{\pi'}$, we assume the same scaling as the SM QCD,
\begin{equation}
    f_{\pi'} \simeq \frac{\sqrt{3}}{2\pi} m_{\rho'} \simeq \frac{1}{g_3}m_O.
\end{equation}
We also assume that $K\simeq 2$, which is a typical value for the production of color-singlet scalars  by gluon fusion (see e.g.~\cite{Ahmed:2016otz}.) Color octets might have different $K$-factors.
Comparing the cross section with the constraints from the di-jet resonance search in~\cite{ATLAS:2019fgd,ATLAS:2025okg}, we obtain $m_O \gtrsim 1.4$ TeV, which is comparable to the constraints from other processes discussed above. Note, however, that the constraint on $m_O$ is sensitive to the relation between $f_{\pi'}$ and $m_O$, which might be different from our naive estimation based on three-flavor QCD, as well as to the $K$-factor. Constraints from other processes by the gauge interaction do not suffer from the former uncertainty.

\subsection{Dark matter}
\label{sec:DM}

In our model, the stable particles are $T$ and the lightest baryon $B$. The stability of $T$ is guaranteed by the unbroken $U(1)_T$ symmetry, while that of $B$ is guaranteed by the unbroken $U(1)_V$ symmetry. Both of them can be dark matter candidates. In the following, we discuss both possibilities. Note that the vector boson $A_D$ can be also stable if $m_{A_D} < 2 m_T$ with vanishing kinetic mixing with the hypercharge gauge boson, but this case is not viable as dark matter, since the abundance of $A_D$ cannot be reduced.


\subsubsection{NGB dark matter}

For $m_{T} < m_{A_D}$, the freeze-out abundance of $T$ is determined by the annihilation of $T$ involving the kinetic mixing between $U(1)_D$ and $U(1)_Y$. The required kinetic mixing is excluded by direct-detection experiments unless $m_{T} < 10$ GeV~\cite{Pospelov:2007mp}, which requires small $\Lambda'$ and is excluded by collider search for the octet NGBs (see Sec.~\ref{sec:collider}.)
For $m_{T} > m_{A_D}$, the freeze-out abundance of $T$ is determined by the annihilation into a pair of $A_D$ whose cross section is given by~\cite{Contino:2020god}
\begin{align}
     \sigma v =&  \frac{e_D^4 (1+a)^4}{4\pi m_T^2}\ 
    F\biggl(\frac{m_{A_D}^2}{m_T^2}\biggr),  \, \nonumber\\
    \label{eq:F}
    F(x)\equiv& \sqrt{ 1- x} \left( \frac{x^2}{2(x-2)^2} + \frac{(x-2)^2}{2x^2} (K-1)^2 + K^2 + K-1\right),~~ K = \frac{4a}{(1+a)^2}.
\end{align}
The resultant freeze-out abundance of $T$ is
\begin{align}
\label{eq:DM abundance}
      \frac{\rho_T}{s} \simeq 0.4~{\rm eV} \left( \frac{m_{T}}{1.1~{\rm TeV}} \right)^2 
    \left( \frac{0.5}{e_D(1+a)} \right)^4 \frac{1}{F\bigl(m_{A_D}^2/m_T^2\bigr)} \, ,
\end{align}
while the observed dark matter abundance is $\rho_{\rm DM}/s \simeq 0.4$ eV.
From Eq.~\eqref{eq:DM abundance} and the relation between $\Lambda'$ and $m_{T}$, we obtain
\begin{align}
\label{eq:relation}
    \Lambda' &\simeq 0.7~{\rm TeV} \left(\frac{m_{T}}{100~{\rm GeV}}\right)^{1/2}\frac{1+a}{a^{1/2}} F^{1/4} \simeq 2.5~{\rm TeV} \frac{(1+a)^2}{a^{1/2}} \frac{e_D}{0.5} F^{1/2}, \\
    \label{eq:eD}
    e_D &\simeq 0.15 \frac{1}{1+a}  \left(\frac{m_{T}}{100~{\rm GeV}}\right)^{1/2} \frac{1}{F^{1/4}}.
\end{align}

$m_{T} < 10$ GeV is excluded by the CMB constraint~\cite{Kawasaki:2021etm} and $m_{T} < 100$ GeV is disfavored by the Fermi-LAT constraint~\cite{Fermi-LAT:2015att}. From Eqs.~\eqref{eq:mAD}, \eqref{eq:mNGB}, \eqref{eq:F}, and \eqref{eq:relation}, we find
\begin{equation}
\label{eq:Llow}
    \Lambda' \gtrsim 1~{\rm TeV},
\end{equation}
unless $m_{A_D}\simeq m_T$, for which $F\ll 1$ and smaller $\Lambda'$ is allowed.
The bound corresponds to $v' > 3\times 10^{12}$ GeV.
This lower bound on $\Lambda'$ has uncertainty from that of the relation between $\Lambda'$ and $m_{T}$. The uncertainty can be removed by expressing it in terms of the lower bound on $m_{O}$,
\begin{equation}
\label{eq:octetlow}
    m_O > 0.8~{\rm TeV},
\end{equation}
which is weaker than the collider bound.

$T$ scatters with nucleon via the kinetic mixing $\epsilon$ between $A_D$ and the photon. The scattering cross section between $T$ and a nucleus per nucleon is~\cite{Co:2016akw}
\begin{equation}
\sigma_n = \frac{\epsilon^2 (1+a)^2 e_D^2 e^2  m_N^2}{\pi m_{A_D}^4} \left(\frac{Z}{A}\right)^2 \simeq 6\times 10^{-49}~{\rm cm}^2 \frac{m_{T}}{100~{\rm GeV}} \left( \frac{50~{\rm GeV}}{m_{A_D}}\right)^4 \left(\frac{\epsilon}{10^{-5}}\right)^2 \left(\frac{Z/A}{54/131}\right)^2 \frac{1}{F^{1/2}},
\end{equation}
where in the second equality we used Eq.~\eqref{eq:eD}.
The cross section depends on the unknown kinetic mixing and cannot be predicted, but one may check the consistency of direct-detection signals with collider signals that depend on the kinetic mixing; see Sec.~\ref{sec:collider}. For example, if dark matter is directly detected by near-future experiments, the kinetic mixing should be large enough that $A_D$ decays promptly at the LHC, and the signal of $O_{T_0}$ is multi-jets plus leptons.

$T$ interacts with gluons via the octet NGB loops~\cite{Contino:2020god}. The interaction of $T$ with the octet NGB includes $\partial T \partial T^* O O/f_{\pi'}^2$. After integrating out the octets, we obtain
\begin{equation}
\frac{g_3^2}{M_c^4} \partial^\mu T \partial_\mu T^* G_{\rho \sigma}^a G^{a,\rho \sigma},~~\frac{1}{M_c^4}\sim \frac{3}{64\pi^2} \frac{1}{f_{\pi'}^2 m_O^2} \simeq  \frac{0.01}{\Lambda^{'4}}.
\end{equation}
The trace anomaly relation~\cite{Shifman:1978zn} gives
\begin{equation}
    2m_N^2 = \langle N| T^\mu_\mu |N\rangle = - \frac{9}{16\pi^2} \langle N| g_3^2 G_{\rho \sigma}^a G^{a,\rho \sigma}|N \rangle + \sum_{q=u,d,s} \langle N| m_q q \bar{q}|N \rangle,
\end{equation}
and the second term on the right hand side is negligible~\cite{Abdel-Rehim:2016won}. The scattering cross section between $T$ and a nucleus per nucleon is
\begin{equation}
    \sigma_n \simeq \frac{1}{16\pi} \left(\frac{32\pi^2}{9}\right)^2 \frac{m_T^2 m_N^4}{M_c^8} \simeq 10^{-50}~{\rm cm}^2 \left(\frac{{\rm TeV}}{\Lambda'}\right)^8 \left(\frac{m_T}{100~{\rm GeV}}\right)^2,
\end{equation}
which is below the neutrino floor.

\subsubsection{Baryon dark matter}
\label{sec:baryon DM}

$B$ annihilates via the mirror strong interaction. In order for the freeze-out abundance to explain the observed dark-matter abundance, $m_B$ needs to be ${\cal O} (100)$ TeV, which requires $\Lambda'=$ few ten TeV. The corresponding $v'$ is above $10^{16}$ GeV, for which the correction to the strong CP phase in Eq.~\eqref{eq:dtheta} is too large unless the cutoff scale is much above the gravity scale.

However, even for $\Lambda'= {\cal O} (1)$ TeV, the following process can produce $B$ as the dominant component of dark matter. Assuming that the temperature of the Universe is above $m_{u'}$, the freeze-out abundance of $u'$ exceeds the observed dark matter abundance. With the interactions in Eqs.~\eqref{eq:decay1} and \eqref{eq:decay2}, $u'$ decays and and produces $B$.
The abundance of $B$ right after the production is still too large, but they
 annihilate with each other and the resultant abundance is
\begin{equation}
    \frac{\rho_B}{s} \simeq \left.\frac{m_B H}{\sigma_B v s}\right|_{u'~{\rm decay}} =  0.4~{\rm eV}\left(\frac{m_B}{10~{\rm TeV}} \right)^3 \frac{\left(4\pi/m_B\right)^2}{\sigma_B v} \frac{0.5~{\rm GeV}}{T_{\rm dec}},
\end{equation}
where $T_{\rm dec}$ is given by Eq.~\eqref{eq:Tdec}.
For $\Lambda'= {\cal O} (1)$ TeV, if $u'$ decays around $T=1$ GeV, the observed dark matter abundance can be explained from $B$.
$T$ is also produced by the decay of $u'$ or the annihilation of $B$. Its abundance is
\begin{equation}
    \frac{\rho_T}{s} \simeq \left.\frac{m_T H}{\sigma_T v s}\right|_{u'~{\rm decay}} =  0.4~{\rm eV} \frac{0.5}{e_D} \frac{a^{3/2}/(1+a)^4}{0.07} \frac{\sigma_B v}{0.2 (4\pi/m_B)^2 } \frac{1}{F},
\end{equation}
where in the second equality we fixed $T_{\rm dec}$ so that $B$ is the dominant component of dark matter and we also used the relation between $m_T$, $m_{\rho'}$, and $m_B$. For $\sigma_B v $ smaller than naive expectation $ (4\pi/m_B)^2$ and/or sufficiently small $a$, $T$ is not overproduced as dark matter. For $a=0$, $T$ is massless and does not contribute to dark matter density. 

In the present Universe, $B$ annihilate into NGBs with a cross section
\begin{equation}
\label{eq:B anni}
    \sigma_B v \sim \frac{(4\pi)^2}{m_B^2} \sim 10^{-23}~{\rm cm}^3{\rm s}^{-1} \left(\frac{10~{\rm TeV}}{m_B}\right)^2.
\end{equation}
The NGBs decay into gluons, which subsequently produce gamma rays.
The cross section is just around the upper bound set by the Fermi-LAT~\cite{Fermi-LAT:2015att}, so depending on the exact value of the cross section, $B$ dark matter may be already excluded. Even if the cross section is smaller than the reference value in Eq.~\eqref{eq:B anni} by a factor of ${\cal O}(1-10)$, the observation of the center of the galaxy by the CTAO can discover the indirect signal of $B$ dark matter~\cite{CTA:2020qlo}.

Because $B$ is composed of colored particles, it should interact with gluons via higher dimensional operators,
\begin{equation}
    \frac{4\pi c_B g_3^2}{m_{\rho'}^3} B \bar{B} G_{\rho \sigma}^a G^{a,\rho \sigma},
\end{equation}
where we determined the coefficient via the naive dimensional analysis~\cite{Manohar:1983md,Georgi:1992dw} and $c_B$ is expected to be ${\cal O} (1)$.
The scattering cross section between $B$ and a nucleus per nucleon is
\begin{equation}
    \sigma_n \simeq \frac{(4\pi c_B)^2}{16\pi} \left(\frac{32\pi^2}{9}\right)^2 \frac{ m_N^4}{m_{\rho'}^6} \simeq 5 \times 10^{-48}~{\rm cm}^2 \left(\frac{10~{\rm TeV}}{m_B}\right)^6 c_B^2,
\end{equation}
where we used $m_B = m_{\rho'} \times 940/770$. In the parameter region that satisfies the collider bound $m_O > 1.5$ TeV, the predicted cross section is below the neutrino floor for $c_B=1$. If $c_B >1$, it may be possible to directly detect $B$ dark matter.

\subsection{Baryon asymmetry by leptogenesis}

The baryon asymmetry of the universe can be explained by leptogenesis~\cite{Fukugita:1986hr} from the decay of mirror neutrinos~\cite{Carrasco-Martinez:2023nit}. 
We discuss thermal and non-thermal leptogenesis and the compatibility of the scenarios with our model.

The neutrino mass is given by the following dimension-5 operators,
\begin{equation}
    \frac{a_i}{2M_L} \ell_i \ell_i H H + \frac{a_i}{2M_L} \ell'_i \ell'_i H' H' + \frac{b_{ij}}{M_L} \ell_i \ell'_j H H'.
\end{equation}
After the mirror electroweak symmetry breaking, the mirror neutrinos obtain Majorana masses from the second term  and yukawa couplings to the SM lepton doublets and Higgs from the third term. The SM neutrino mass is determined by the first term and the seesaw contribution from the mirror neutrinos.
To avoid too large SM neutrino masses requires $a_i v^2 /M_L < m_\nu$, where $m_\nu = O(10-100)$ meV is the observed SM neutrino mass.  

\subsubsection{Thermal leptogenesis}
In thermal leptogenesis, the reheating temperature and the mass of the mirror neutrino whose decay produces lepton asymmetry need to be larger than $10^9$ GeV~\cite{Giudice:2003jh,Buchmuller:2004nz}. To satisfy the upper bound on $a_i/M_L$ requires
\begin{equation}
    v' > 8\times 10^{11}~{\rm GeV}  \left(\frac{50~{\rm meV}}{m_\nu}\right)^{1/2},
\end{equation}
which is satisfied in the viable parameter space of our model.
Because the reheating temperature is above $10^9$ GeV, $u'$ and $d'$ are thermalized. Then we need to introduce the interactions in Eqs.~\eqref{eq:decay1} or \eqref{eq:decay2} to let $u'$ and $
d'$ decay.%
\footnote{If two mirror neutrinos are degenerate in their masses, lepton asymmetry can be enhanced~\cite{Covi:1996wh}. $O(1)$ \% degeneracy can lower the required reheating temperature down to $10^{7}$ GeV, so that the production of $u'$ and $d'$ can be avoided. Such a scenario is consistent with our model without introducing the interactions in Eqs.~\eqref{eq:decay1} and \eqref{eq:decay2}.}

\subsubsection{Non-thermal leptogenesis}

If the inflaton (or any particle that dominates the universe) directly decays into the mirror neutrinos, the reheating temperature may be as low as $10^6$ GeV~\cite{Asaka:1999yd}. Then $u'$ and $d'$ are not produced in the early universe, and the interactions in Eqs.~\eqref{eq:decay1} or \eqref{eq:decay2} are not necessary.%
\footnote{
As long as the inflaton mass is below $m_{u'}$, the production of $u'$ by the scattering during thermalization~\cite{Harigaya:2014waa,Harigaya:2016vda,Harigaya:2019tzu,Drees:2022vvn,Mukaida:2022bbo} can be also avoided. 
}

\subsection{Gravitational waves from mirror QCD phase transition}

The mirror QCD dynamics is that of $SU(3)$ with 6 flavors, which is expected to exhibit first-order phase transition~\cite{Pisarski:1983ms}. Gravitational waves are produced during the mirror QCD phase transition. The typical frequency of the gravitational waves is
\begin{equation}
    f\sim 1 {\rm mHz}~\frac{\Lambda'}{1~{\rm TeV}} \frac{\beta/H}{100},
\end{equation}
where $(\beta/H)^{-1}$ is the duration of the phase transition in comparison with the cosmological-expansion time scale.

The magnitude of the gravitational waves is highly sensitive to the detail of the phase transition, such as the latent heat $\rho_{\rm lat}$, the energy density of the mirror QCD fluid $\rho_{{\rm QCD}'}$, and the velocity of the bubble wall $v_w$. As long as $\rho_{\rm lat}$ is not much smaller than the total energy density of the universe, the magnitude of gravitational waves around the peak frequency is dominated by bubble collisions and is given by~\cite{Caprini:2010xv}
\begin{equation}
    \Omega_{\rm GW} h^2 \simeq 10^{-11} \left( \frac{100}{\beta/H}\right)^2 \left(\frac{v_w}{1/\sqrt{3}}\right)^3 \left( \frac{\rho_{\rm lat}}{\rho_{{\rm QCD}}'}\right)^2.
\end{equation}
The gravitational waves with $f \sim 1$ mHz can be detected by LISA if $\Omega_{\rm GW} h^2\gtrsim 10^{-11}$~\cite{Schmitz:2020syl}.

\section{Summary}
\label{sec:summary}

We propose a solution to the strong CP problem in which a Peccei-Quinn (PQ) symmetry arises accidentally from a chiral $U(1)$ gauge symmetry. A $Z_2$ symmetry exchanging Standard-Model particles with their mirror counterparts is imposed. The PQ symmetry is both spontaneously and explicitly broken by mirror QCD dynamics, and the theory contains no massless colored fermions or a light QCD axion.

The model predicts 35 Nambu-Goldstone bosons (NGBs). One is eaten by the \(U(1)\) gauge boson, 32 organize into four color octets, and two are neutral under the Standard Model gauge group and are stable. Two of the octet NGBs decay into two gluons and a stable NGB or a gluon, a $U(1)$ gauge boson, and a stable NGB, leading to collider signatures of jets, leptons, and missing energy at the LHC. Another octet state decays into a gluon and a \(U(1)\) gauge boson. Depending on the lifetime of the \(U(1)\) gauge boson, this process gives rise to signatures such as  jets plus missing energy, jets plus displaced vertices, or multijet events with leptons. The remaining octet state decays into a pair of gluons, resulting in four-jet final states with two dijet resonances; it can also be singly produced, leading to dijet resonance signatures.

The stable NGBs can be massive and serve as dark matter candidates. The gauge boson of the chiral $U(1)$ gauge symmetry provides a vector portal connecting the Standard-Model particles with dark matter. Indirect detection experiments constrain their mass to be above 100~GeV, which in turn implies that the octet masses exceed approximately \(1\)~TeV. If instead the stable NGBs are massless, they contribute to the radiation energy density of the Universe with \(\Delta N_{\rm eff} \simeq 0.054\).

Unlike most solutions based on accidental PQ symmetry, the model does not suffer from stable domain walls. Metastable domain walls may form during the mirror QCD phase transition, but they decay via higher-dimensional operators. The absence of stable domain walls allows the reheating temperature to exceed the mirror confinement scale, enabling leptogenesis via mirror neutrinos. Moreover, the energy fraction stored in domain walls at the time of their decay can be sufficiently large to generate an observable stochastic gravitational-wave background.
Also, the mirror QCD phase transition is expected to be of first order and produce gravitational waves with frequencies around mHz.

As we have shown, the model yields a rich set of experimental signatures across multiple frontiers, including collider searches, indirect-detection experiments, measurements of neutron and proton electric dipole moments, observations of the cosmic microwave background, and gravitational-wave detectors. Our work motivates a synergistic experimental program to probe the origin of the small CP violation in the strong interaction.

\acknowledgments
KH thanks Liantao Wang for useful discussion.
This work was supported by JSPS KAKENHI Grant Nos.\ 24K17042 [HF], 25H00638 [HF], 26H00403 [HF], 26K17131 [HF], the Department of Energy grant DE-SC0009924 [KH], and the World Premier International Research Center Initiative (WPI), MEXT, Japan (Kavli IPMU) [KH]. 
In this research work, HF used the UTokyo Azure (\url{https://utelecon.adm.u-tokyo.ac.jp/en/research_computing/utokyo_azure/}).


\appendix

\section{Unbroken discrete symmetries and decay of NGBs}
\label{app:discrete symmetry}

In this appendix, we introduce (approximate) discrete symmetries of the $SU(3)_c\times SU(3)_c'\times U(1)_D$ gauge interaction and discuss the decay modes of NGBs.

\subsection{Discrete symmetries}

\begin{itemize}
    \item 
    $C_D$
is a charge-conjugation symmetry that only conjugates $U(1)_D$ charge, 
\begin{align}
    \psi_u \rightarrow \psi_d,~~ \psi_d \rightarrow \psi_u,~~
    \bar{\psi}_u \rightarrow \bar{\psi}_d, ~~\bar{\psi}_d \rightarrow \bar{\psi}_u,~~ A_D^\mu \rightarrow - A_D^\mu,~~G_\mu^a \rightarrow G_\mu^a.
\end{align}
The transformation laws of the NGBs under $C_D$ can be derived by computing those of the corresponding currents. We find
\begin{equation}
    O \rightarrow O,~~O_{T_0} \rightarrow - O_{T_0}. 
\end{equation}
This symmetry is explicitly broken by the kinetic mixing between the $U(1)_D$ and $U(1)_Y$ gauge fields and the absence of charge-conjugation symmetry of $U(1)_Y$, but its effect on the $SU(3)_c'$ dynamics is suppressed by the smallness of the kinetic mixing and the necessity of SM fermion loops.
\item 
$C_cP$ is a combination of the CP symmetry and $C_D$ that only conjugates the $SU(3)_c\times SU(3)_c'$ charges,
\begin{align}
    \psi_u(t,{\bf x}) \rightarrow i \sigma^2 \psi_d^*(t,-{\bf x}),~~ \psi_d(t,{\bf x}) \rightarrow -i \sigma^2 \psi_u^*(t,-{\bf x}), \\
    \bar{\psi}_u(t,{\bf x}) \rightarrow i \sigma^2 \bar{\psi}_d^*(t,-{\bf x}),~~ \bar{\psi}_d(t,{\bf x}) \rightarrow -i \sigma^2 \bar{\psi}_u^*(t,-{\bf x}), \\
    A_{D,\mu} (t,{\bf x}) \rightarrow  A_{D}^\mu (t,-{\bf x}),~~\lambda^a G_\mu^a(t,{\bf x}) \rightarrow - \lambda^{a*} G^{a,\mu}(t,-{\bf x}),
\end{align}
where $i \sigma^2$ acts on the spinor indices. Under this symmetry, NGBs transform as
\begin{align}
    T(t,{\bf x}) \rightarrow T(t,-{\bf x}),~~\lambda^aO_T^a(t,{\bf x}) \rightarrow  \lambda^{a*}O_T^a(t,-{\bf x}), \nonumber \\
    \lambda^aO^a(t,{\bf x}) \rightarrow - \lambda^{a*}O^a(t,-{\bf x}), ~~ \lambda^aO_{T_0}^a(t,{\bf x}) \rightarrow  \lambda^{a*}O_{T_0}^a(t,-{\bf x}).
\end{align}
\item 
$C_c$
is an approximate symmetry for $e_D=0$,
\begin{equation}
    \psi_u \rightarrow \bar{\psi}_d,~~\psi_d \rightarrow - \bar{\psi}_u,~~G^a \lambda^a \rightarrow - G^a \lambda^{a*}.
\end{equation}
The NGBs transform as
\begin{equation}
    T \rightarrow -T,~~T_0 \rightarrow - T_0 ~~\lambda^aO^a \rightarrow \lambda^{a*}O^a,~~ \lambda^aO_T^a \rightarrow -\lambda^{a*}O_T^a,~~ \lambda^aO_{T_0}^a \rightarrow -\lambda^{a*}O_{T_0}^a.
\end{equation}
Here we include the transformation law of the would-be NGB $T_0$, which can be used to constrain the decay involving the longitudinal mode of $A_D$.

\end{itemize}

\subsection{Decay of NGBs}

Based on the discrete symmetries introduced above, we discuss the decay of NGBs.

\begin{itemize}
\item 
$O$:
No symmetry forbids $O\rightarrow gg$. In fact, the decay occurs through the interaction in Eq.~\eqref{eq:WZW}. $O\rightarrow g A_D$ is forbidden by $C_D$. The dominant decay mode is thus $O\rightarrow gg$.
\item 
$O_{T_0}$:
$C_D$ forbids the decay of $O_{T_0}$ into a pair of gluons. No symmetry forbids the decay of $O_{T_0}$ into $A_Dg$, but $C_cP$ forbids the interaction of $O_{T_0}$ with $A_D g$ without the Levi-Civita tensor, so the interaction should be of the form $O_{T_0} F_D^{\mu \nu} \tilde{G}_{\mu \nu}$, which indeed arises through the ABJ anomaly of the symmetry associated with $O_{T_0}$. The dominant decay mode is $O_{T_0} \rightarrow A_D g$. 
\item 
$O_{T}$:
Because of $U(1)_T$ symmetry, the decay product of $O_T$ includes $T$. 
$C_c$ forbids $O_T \rightarrow T g$, so this decay needs quantum corrections from $A_D$ and is suppressed by $e_D^4/(16\pi^2)^2$. $O_T \rightarrow T gg$ is not forbidden by any symmetry and is suppressed by $g_3^4/(16\pi^2)$. $O_T \rightarrow T A_Dg$ is also not forbidden by any symmetry, but $C_cP$ requires that the interaction responsible for this decay involves the Levi-Civita tensor.
Indeed, the decay comes from the Wess-Zumino-Witten term corresponding to the anomaly of the form $\partial J_A \supset \epsilon_{\mu \nu \rho \sigma} A_{\mu} A_\nu G_{\rho \sigma}$, where $A$ is the (non-Abelian) background gauge field associated with the axial symmetry\,\cite{Witten:1983tw}. This means that $A_D$ in the final state should be understood as the would-be NGB $T_0$, so the rate
is suppressed only by $g_3^2/(16\pi^2)$ without suppression by $e_D^2$. Note that the decay $O_T \rightarrow T T_0 g$ is consistent with $C_c$.
We conclude that the dominant decay mode is $O_T \rightarrow T gg$ and $O_T \rightarrow T A_Dg$.
\footnote{In the setup in~\cite{Contino:2020god}, where $SU(3)_c$ is replaced with $SU(2)_w$, the decay corresponding to $O_T\rightarrow Tgg$ ($3_\pm \rightarrow 1_\pm VV$) is suppressed because of the approximate $G$ parity that is intrinsic to $SU(2)_w$. The reference also argues that the decay corresponding to $O_T \rightarrow T A_Dg$ ($3_\pm \rightarrow 1_\pm A_D V$) is suppressed, which we could not confirm. This cannot be explained by the $G$ parity; in the NGB picture, $T_0$ ($1_0$) has an odd $G$ parity, and $3_\pm \rightarrow 1_\pm 1_0 V$ preserves the $G$ parity.}

\end{itemize}

\bibliographystyle{JHEP}

\bibliography{papers}

\providecommand{\href}[2]{#2}\begingroup\raggedright\begin{thebibliography}{100}

\bibitem{Baker:2006ts}
C.A.~Baker et~al., \emph{{An Improved experimental limit on the electric dipole
  moment of the neutron}},
  \href{https://doi.org/10.1103/PhysRevLett.97.131801}{\emph{Phys. Rev. Lett.}
  {\bfseries 97} (2006) 131801}
  [\href{https://arxiv.org/abs/hep-ex/0602020}{{\ttfamily hep-ex/0602020}}].

\bibitem{Pospelov:1999ha}
M.~Pospelov and A.~Ritz, \emph{{Theta induced electric dipole moment of the
  neutron via QCD sum rules}},
  \href{https://doi.org/10.1103/PhysRevLett.83.2526}{\emph{Phys. Rev. Lett.}
  {\bfseries 83} (1999) 2526}
  [\href{https://arxiv.org/abs/hep-ph/9904483}{{\ttfamily hep-ph/9904483}}].

\bibitem{Peccei:1977hh}
R.D.~Peccei and H.R.~Quinn, \emph{{CP Conservation in the Presence of
  Instantons}}, \href{https://doi.org/10.1103/PhysRevLett.38.1440}{\emph{Phys.
  Rev. Lett.} {\bfseries 38} (1977) 1440}.

\bibitem{Peccei:1977ur}
R.D.~Peccei and H.R.~Quinn, \emph{{Constraints Imposed by CP Conservation in
  the Presence of Instantons}},
  \href{https://doi.org/10.1103/PhysRevD.16.1791}{\emph{Phys. Rev. D}
  {\bfseries 16} (1977) 1791}.

\bibitem{Weinberg:1977ma}
S.~Weinberg, \emph{{A New Light Boson?}},
  \href{https://doi.org/10.1103/PhysRevLett.40.223}{\emph{Phys. Rev. Lett.}
  {\bfseries 40} (1978) 223}.

\bibitem{Wilczek:1977pj}
F.~Wilczek, \emph{{Problem of Strong $P$ and $T$ Invariance in the Presence of
  Instantons}}, \href{https://doi.org/10.1103/PhysRevLett.40.279}{\emph{Phys.
  Rev. Lett.} {\bfseries 40} (1978) 279}.

\bibitem{Holman:1992us}
R.~Holman, S.D.H.~Hsu, T.W.~Kephart, E.W.~Kolb, R.~Watkins and L.M.~Widrow,
  \emph{{Solutions to the strong CP problem in a world with gravity}},
  \href{https://doi.org/10.1016/0370-2693(92)90491-L}{\emph{Phys. Lett. B}
  {\bfseries 282} (1992) 132}
  [\href{https://arxiv.org/abs/hep-ph/9203206}{{\ttfamily hep-ph/9203206}}].

\bibitem{Barr:1992qq}
S.M.~Barr and D.~Seckel, \emph{{Planck scale corrections to axion models}},
  \href{https://doi.org/10.1103/PhysRevD.46.539}{\emph{Phys. Rev. D} {\bfseries
  46} (1992) 539}.

\bibitem{Kamionkowski:1992mf}
M.~Kamionkowski and J.~March-Russell, \emph{{Planck scale physics and the
  Peccei-Quinn mechanism}},
  \href{https://doi.org/10.1016/0370-2693(92)90492-M}{\emph{Phys. Lett. B}
  {\bfseries 282} (1992) 137}
  [\href{https://arxiv.org/abs/hep-th/9202003}{{\ttfamily hep-th/9202003}}].

\bibitem{Dine:1992vx}
M.~Dine, \emph{{Problems of naturalness: Some lessons from string theory}},  in
  \emph{{Conference on Topics in Quantum Gravity}}, 7, 1992
  [\href{https://arxiv.org/abs/hep-th/9207045}{{\ttfamily hep-th/9207045}}].

\bibitem{Kallosh:1995hi}
R.~Kallosh, A.D.~Linde, D.A.~Linde and L.~Susskind, \emph{{Gravity and global
  symmetries}}, \href{https://doi.org/10.1103/PhysRevD.52.912}{\emph{Phys. Rev.
  D} {\bfseries 52} (1995) 912}
  [\href{https://arxiv.org/abs/hep-th/9502069}{{\ttfamily hep-th/9502069}}].

\bibitem{Arkani-Hamed:2006emk}
N.~Arkani-Hamed, L.~Motl, A.~Nicolis and C.~Vafa, \emph{{The String landscape,
  black holes and gravity as the weakest force}},
  \href{https://doi.org/10.1088/1126-6708/2007/06/060}{\emph{JHEP} {\bfseries
  06} (2007) 060} [\href{https://arxiv.org/abs/hep-th/0601001}{{\ttfamily
  hep-th/0601001}}].

\bibitem{Banks:2010zn}
T.~Banks and N.~Seiberg, \emph{{Symmetries and Strings in Field Theory and
  Gravity}}, \href{https://doi.org/10.1103/PhysRevD.83.084019}{\emph{Phys. Rev.
  D} {\bfseries 83} (2011) 084019}
  [\href{https://arxiv.org/abs/1011.5120}{{\ttfamily 1011.5120}}].

\bibitem{Harlow:2018tng}
D.~Harlow and H.~Ooguri, \emph{{Symmetries in quantum field theory and quantum
  gravity}}, \href{https://doi.org/10.1007/s00220-021-04040-y}{\emph{Commun.
  Math. Phys.} {\bfseries 383} (2021) 1669}
  [\href{https://arxiv.org/abs/1810.05338}{{\ttfamily 1810.05338}}].

\bibitem{Georgi:1981pu}
H.M.~Georgi, L.J.~Hall and M.B.~Wise, \emph{{Grand Unified Models With an
  Automatic {Peccei-Quinn} Symmetry}},
  \href{https://doi.org/10.1016/0550-3213(81)90433-8}{\emph{Nucl. Phys. B}
  {\bfseries 192} (1981) 409}.

\bibitem{Lazarides:1985bj}
G.~Lazarides, C.~Panagiotakopoulos and Q.~Shafi, \emph{{Phenomenology and
  Cosmology With Superstrings}},
  \href{https://doi.org/10.1103/PhysRevLett.56.432}{\emph{Phys. Rev. Lett.}
  {\bfseries 56} (1986) 432}.

\bibitem{Randall:1992ut}
L.~Randall, \emph{{Composite axion models and Planck scale physics}},
  \href{https://doi.org/10.1016/0370-2693(92)91928-3}{\emph{Phys. Lett. B}
  {\bfseries 284} (1992) 77}.

\bibitem{Dias:2002hz}
A.G.~Dias, V.~Pleitez and M.D.~Tonasse, \emph{{Naturally light invisible axion
  and local Z(13) x Z(3) symmetries}},
  \href{https://doi.org/10.1103/PhysRevD.69.015007}{\emph{Phys. Rev. D}
  {\bfseries 69} (2004) 015007}
  [\href{https://arxiv.org/abs/hep-ph/0210172}{{\ttfamily hep-ph/0210172}}].

\bibitem{Dias:2002gg}
A.G.~Dias, V.~Pleitez and M.D.~Tonasse, \emph{{Naturally light invisible axion
  in models with large local discrete symmetries}},
  \href{https://doi.org/10.1103/PhysRevD.67.095008}{\emph{Phys. Rev. D}
  {\bfseries 67} (2003) 095008}
  [\href{https://arxiv.org/abs/hep-ph/0211107}{{\ttfamily hep-ph/0211107}}].

\bibitem{Babu:2002ic}
K.S.~Babu, I.~Gogoladze and K.~Wang, \emph{{Stabilizing the axion by discrete
  gauge symmetries}},
  \href{https://doi.org/10.1016/S0370-2693(03)00411-8}{\emph{Phys. Lett. B}
  {\bfseries 560} (2003) 214}
  [\href{https://arxiv.org/abs/hep-ph/0212339}{{\ttfamily hep-ph/0212339}}].

\bibitem{Choi:2009jt}
K.-S.~Choi, H.P.~Nilles, S.~Ramos-Sanchez and P.K.S.~Vaudrevange,
  \emph{{Accions}},
  \href{https://doi.org/10.1016/j.physletb.2009.04.028}{\emph{Phys. Lett. B}
  {\bfseries 675} (2009) 381}
  [\href{https://arxiv.org/abs/0902.3070}{{\ttfamily 0902.3070}}].

\bibitem{Carpenter:2009zs}
L.M.~Carpenter, M.~Dine and G.~Festuccia, \emph{{Dynamics of the Peccei Quinn
  Scale}}, \href{https://doi.org/10.1103/PhysRevD.80.125017}{\emph{Phys. Rev.
  D} {\bfseries 80} (2009) 125017}
  [\href{https://arxiv.org/abs/0906.1273}{{\ttfamily 0906.1273}}].

\bibitem{Harigaya:2013vja}
K.~Harigaya, M.~Ibe, K.~Schmitz and T.T.~Yanagida, \emph{{Peccei-Quinn symmetry
  from a gauged discrete R symmetry}},
  \href{https://doi.org/10.1103/PhysRevD.88.075022}{\emph{Phys. Rev. D}
  {\bfseries 88} (2013) 075022}
  [\href{https://arxiv.org/abs/1308.1227}{{\ttfamily 1308.1227}}].

\bibitem{Harigaya:2015soa}
K.~Harigaya, M.~Ibe, K.~Schmitz and T.T.~Yanagida, \emph{{Peccei-Quinn Symmetry
  from Dynamical Supersymmetry Breaking}},
  \href{https://doi.org/10.1103/PhysRevD.92.075003}{\emph{Phys. Rev. D}
  {\bfseries 92} (2015) 075003}
  [\href{https://arxiv.org/abs/1505.07388}{{\ttfamily 1505.07388}}].

\bibitem{Fukuda:2017ylt}
H.~Fukuda, M.~Ibe, M.~Suzuki and T.T.~Yanagida, \emph{{A ''gauged'' $U(1)$
  Peccei{\textendash}Quinn symmetry}},
  \href{https://doi.org/10.1016/j.physletb.2017.05.071}{\emph{Phys. Lett. B}
  {\bfseries 771} (2017) 327}
  [\href{https://arxiv.org/abs/1703.01112}{{\ttfamily 1703.01112}}].

\bibitem{DiLuzio:2017tjx}
L.~Di~Luzio, E.~Nardi and L.~Ubaldi, \emph{{Accidental Peccei-Quinn symmetry
  protected to arbitrary order}},
  \href{https://doi.org/10.1103/PhysRevLett.119.011801}{\emph{Phys. Rev. Lett.}
  {\bfseries 119} (2017) 011801}
  [\href{https://arxiv.org/abs/1704.01122}{{\ttfamily 1704.01122}}].

\bibitem{Duerr:2017amf}
M.~Duerr, K.~Schmidt-Hoberg and J.~Unwin, \emph{{Protecting the Axion with
  Local Baryon Number}},
  \href{https://doi.org/10.1016/j.physletb.2018.03.054}{\emph{Phys. Lett. B}
  {\bfseries 780} (2018) 553}
  [\href{https://arxiv.org/abs/1712.01841}{{\ttfamily 1712.01841}}].

\bibitem{Ibe:2018hir}
M.~Ibe, M.~Suzuki and T.T.~Yanagida, \emph{{$B-L$ as a Gauged Peccei-Quinn
  Symmetry}}, \href{https://doi.org/10.1007/JHEP08(2018)049}{\emph{JHEP}
  {\bfseries 08} (2018) 049}
  [\href{https://arxiv.org/abs/1805.10029}{{\ttfamily 1805.10029}}].

\bibitem{Lillard:2018fdt}
B.~Lillard and T.M.P.~Tait, \emph{{A High Quality Composite Axion}},
  \href{https://doi.org/10.1007/JHEP11(2018)199}{\emph{JHEP} {\bfseries 11}
  (2018) 199} [\href{https://arxiv.org/abs/1811.03089}{{\ttfamily
  1811.03089}}].

\bibitem{Fukuda:2018oco}
H.~Fukuda, M.~Ibe, M.~Suzuki and T.T.~Yanagida, \emph{{Gauged Peccei-Quinn
  symmetry {\textemdash} A case of simultaneous breaking of SUSY and PQ
  symmetry}}, \href{https://doi.org/10.1007/JHEP07(2018)128}{\emph{JHEP}
  {\bfseries 07} (2018) 128}
  [\href{https://arxiv.org/abs/1803.00759}{{\ttfamily 1803.00759}}].

\bibitem{Nakai:2021nyf}
Y.~Nakai and M.~Suzuki, \emph{{Axion Quality from Superconformal Dynamics}},
  \href{https://doi.org/10.1016/j.physletb.2021.136239}{\emph{Phys. Lett. B}
  {\bfseries 816} (2021) 136239}
  [\href{https://arxiv.org/abs/2102.01329}{{\ttfamily 2102.01329}}].

\bibitem{Contino:2021ayn}
R.~Contino, A.~Podo and F.~Revello, \emph{{Chiral models of composite axions
  and accidental Peccei-Quinn symmetry}},
  \href{https://doi.org/10.1007/JHEP04(2022)180}{\emph{JHEP} {\bfseries 04}
  (2022) 180} [\href{https://arxiv.org/abs/2112.09635}{{\ttfamily
  2112.09635}}].

\bibitem{Choi:2022fha}
G.~Choi and T.T.~Yanagida, \emph{{High quality axion in supersymmetric
  models}}, \href{https://doi.org/10.1007/JHEP12(2022)067}{\emph{JHEP}
  {\bfseries 12} (2022) 067}
  [\href{https://arxiv.org/abs/2209.09290}{{\ttfamily 2209.09290}}].

\bibitem{Gherghetta:2025kff}
T.~Gherghetta, H.~Murayama, B.~Noether and P.~Qu{\'\i}lez, \emph{{A
  High-Quality Axion from Exact SUSY Chiral Dynamics}},
  \href{https://arxiv.org/abs/2508.21813}{{\ttfamily 2508.21813}}.

\bibitem{Agrawal:2025mke}
P.~Agrawal, A.~Hook, V.~Loladze and M.~Reig, \emph{{Axion quality problem: keep
  calm and baryon}}, \href{https://doi.org/10.1007/JHEP03(2026)041}{\emph{JHEP}
  {\bfseries 03} (2026) 041}
  [\href{https://arxiv.org/abs/2510.07366}{{\ttfamily 2510.07366}}].

\bibitem{Babu:2026yqp}
K.S.~Babu, S.C.~Chandrasekar and Z.~Tavartkiladze, \emph{{Fermion Mass
  Hierarchy and a High Quality Axion From Gauged U(1) Flavor Symmetry}},
  \href{https://arxiv.org/abs/2602.24253}{{\ttfamily 2602.24253}}.

\bibitem{Choi:1985cb}
K.~Choi and J.E.~Kim, \emph{{DYNAMICAL AXION}},
  \href{https://doi.org/10.1103/PhysRevD.32.1828}{\emph{Phys. Rev. D}
  {\bfseries 32} (1985) 1828}.

\bibitem{Rubakov:1997vp}
V.A.~Rubakov, \emph{{Grand unification and heavy axion}},
  \href{https://doi.org/10.1134/1.567390}{\emph{JETP Lett.} {\bfseries 65}
  (1997) 621} [\href{https://arxiv.org/abs/hep-ph/9703409}{{\ttfamily
  hep-ph/9703409}}].

\bibitem{Berezhiani:2000gh}
Z.~Berezhiani, L.~Gianfagna and M.~Giannotti, \emph{{Strong CP problem and
  mirror world: The Weinberg-Wilczek axion revisited}},
  \href{https://doi.org/10.1016/S0370-2693(00)01392-7}{\emph{Phys. Lett. B}
  {\bfseries 500} (2001) 286}
  [\href{https://arxiv.org/abs/hep-ph/0009290}{{\ttfamily hep-ph/0009290}}].

\bibitem{Hook:2014cda}
A.~Hook, \emph{{Anomalous solutions to the strong CP problem}},
  \href{https://doi.org/10.1103/PhysRevLett.114.141801}{\emph{Phys. Rev. Lett.}
  {\bfseries 114} (2015) 141801}
  [\href{https://arxiv.org/abs/1411.3325}{{\ttfamily 1411.3325}}].

\bibitem{Fukuda:2015ana}
H.~Fukuda, K.~Harigaya, M.~Ibe and T.T.~Yanagida, \emph{{Model of visible QCD
  axion}}, \href{https://doi.org/10.1103/PhysRevD.92.015021}{\emph{Phys. Rev.
  D} {\bfseries 92} (2015) 015021}
  [\href{https://arxiv.org/abs/1504.06084}{{\ttfamily 1504.06084}}].

\bibitem{Dunsky:2023ucb}
D.I.~Dunsky, L.J.~Hall and K.~Harigaya, \emph{{A heavy QCD axion and the mirror
  world}}, \href{https://doi.org/10.1007/JHEP02(2024)212}{\emph{JHEP}
  {\bfseries 02} (2024) 212}
  [\href{https://arxiv.org/abs/2302.04274}{{\ttfamily 2302.04274}}].

\bibitem{Gherghetta:2016fhp}
T.~Gherghetta, N.~Nagata and M.~Shifman, \emph{{A Visible QCD Axion from an
  Enlarged Color Group}},
  \href{https://doi.org/10.1103/PhysRevD.93.115010}{\emph{Phys. Rev. D}
  {\bfseries 93} (2016) 115010}
  [\href{https://arxiv.org/abs/1604.01127}{{\ttfamily 1604.01127}}].

\bibitem{Gaillard:2018xgk}
M.K.~Gaillard, M.B.~Gavela, R.~Houtz, P.~Quilez and R.~Del~Rey, \emph{{Color
  unified dynamical axion}},
  \href{https://doi.org/10.1140/epjc/s10052-018-6396-6}{\emph{Eur. Phys. J. C}
  {\bfseries 78} (2018) 972}
  [\href{https://arxiv.org/abs/1805.06465}{{\ttfamily 1805.06465}}].

\bibitem{Valenti:2022tsc}
A.~Valenti, L.~Vecchi and L.-X.~Xu, \emph{{Grand Color axion}},
  \href{https://doi.org/10.1007/JHEP10(2022)025}{\emph{JHEP} {\bfseries 10}
  (2022) 025} [\href{https://arxiv.org/abs/2206.04077}{{\ttfamily
  2206.04077}}].

\bibitem{Bedi:2024kxe}
R.~Bedi, T.~Gherghetta and K.~Harigaya, \emph{{Solving the strong CP problem
  with massless grand-color quarks}},
  \href{https://doi.org/10.1007/JHEP02(2025)083}{\emph{JHEP} {\bfseries 02}
  (2025) 083} [\href{https://arxiv.org/abs/2408.11246}{{\ttfamily
  2408.11246}}].

\bibitem{Flynn:1987rs}
J.M.~Flynn and L.~Randall, \emph{{A Computation of the Small Instanton
  Contribution to the Axion Potential}},
  \href{https://doi.org/10.1016/0550-3213(87)90089-7}{\emph{Nucl. Phys. B}
  {\bfseries 293} (1987) 731}.

\bibitem{Agrawal:2017ksf}
P.~Agrawal and K.~Howe, \emph{{Factoring the Strong CP Problem}},
  \href{https://doi.org/10.1007/JHEP12(2018)029}{\emph{JHEP} {\bfseries 12}
  (2018) 029} [\href{https://arxiv.org/abs/1710.04213}{{\ttfamily
  1710.04213}}].

\bibitem{Csaki:2019vte}
C.~Cs{\'a}ki, M.~Ruhdorfer and Y.~Shirman, \emph{{UV Sensitivity of the Axion
  Mass from Instantons in Partially Broken Gauge Groups}},
  \href{https://doi.org/10.1007/JHEP04(2020)031}{\emph{JHEP} {\bfseries 04}
  (2020) 031} [\href{https://arxiv.org/abs/1912.02197}{{\ttfamily
  1912.02197}}].

\bibitem{Gherghetta:2020keg}
T.~Gherghetta, V.V.~Khoze, A.~Pomarol and Y.~Shirman, \emph{{The Axion Mass
  from 5D Small Instantons}},
  \href{https://doi.org/10.1007/JHEP03(2020)063}{\emph{JHEP} {\bfseries 03}
  (2020) 063} [\href{https://arxiv.org/abs/2001.05610}{{\ttfamily
  2001.05610}}].

\bibitem{Kitano:2021fdl}
R.~Kitano and W.~Yin, \emph{{Strong CP problem and axion dark matter with small
  instantons}}, \href{https://doi.org/10.1007/JHEP07(2021)078}{\emph{JHEP}
  {\bfseries 07} (2021) 078}
  [\href{https://arxiv.org/abs/2103.08598}{{\ttfamily 2103.08598}}].

\bibitem{Aoki:2024usv}
T.~Aoki, M.~Ibe, S.~Shirai and K.~Watanabe, \emph{{Small instanton effects on
  composite axion mass}},
  \href{https://doi.org/10.1007/JHEP07(2024)269}{\emph{JHEP} {\bfseries 07}
  (2024) 269} [\href{https://arxiv.org/abs/2404.19342}{{\ttfamily
  2404.19342}}].

\bibitem{Harigaya:2016rwr}
K.~Harigaya and Y.~Nomura, \emph{{Light Chiral Dark Sector}},
  \href{https://doi.org/10.1103/PhysRevD.94.035013}{\emph{Phys. Rev. D}
  {\bfseries 94} (2016) 035013}
  [\href{https://arxiv.org/abs/1603.03430}{{\ttfamily 1603.03430}}].

\bibitem{Co:2016akw}
R.T.~Co, K.~Harigaya and Y.~Nomura, \emph{{Chiral Dark Sector}},
  \href{https://doi.org/10.1103/PhysRevLett.118.101801}{\emph{Phys. Rev. Lett.}
  {\bfseries 118} (2017) 101801}
  [\href{https://arxiv.org/abs/1610.03848}{{\ttfamily 1610.03848}}].

\bibitem{Contino:2020god}
R.~Contino, A.~Podo and F.~Revello, \emph{{Composite Dark Matter from
  Strongly-Interacting Chiral Dynamics}},
  \href{https://doi.org/10.1007/JHEP02(2021)091}{\emph{JHEP} {\bfseries 02}
  (2021) 091} [\href{https://arxiv.org/abs/2008.10607}{{\ttfamily
  2008.10607}}].

\bibitem{Ibe:2021gil}
M.~Ibe, S.~Kobayashi and K.~Watanabe, \emph{{Chiral composite asymmetric dark
  matter}}, \href{https://doi.org/10.1007/JHEP07(2021)220}{\emph{JHEP}
  {\bfseries 07} (2021) 220}
  [\href{https://arxiv.org/abs/2105.07642}{{\ttfamily 2105.07642}}].

\bibitem{Blinov:2016kte}
N.~Blinov and A.~Hook, \emph{{Solving the Wrong Hierarchy Problem}},
  \href{https://doi.org/10.1007/JHEP06(2016)176}{\emph{JHEP} {\bfseries 06}
  (2016) 176} [\href{https://arxiv.org/abs/1605.03178}{{\ttfamily
  1605.03178}}].

\bibitem{Hall:2018let}
L.J.~Hall and K.~Harigaya, \emph{{Implications of Higgs Discovery for the
  Strong CP Problem and Unification}},
  \href{https://doi.org/10.1007/JHEP10(2018)130}{\emph{JHEP} {\bfseries 10}
  (2018) 130} [\href{https://arxiv.org/abs/1803.08119}{{\ttfamily
  1803.08119}}].

\bibitem{Witten:1980sp}
E.~Witten, \emph{{Large N Chiral Dynamics}},
  \href{https://doi.org/10.1016/0003-4916(80)90325-5}{\emph{Annals Phys.}
  {\bfseries 128} (1980) 363}.

\bibitem{ATLAS:2017jnp}
{\scshape ATLAS} collaboration, \emph{{A search for pair-produced resonances in
  four-jet final states at $\sqrt{s} =$ 13 TeV with the ATLAS detector}},
  \href{https://doi.org/10.1140/epjc/s10052-018-5693-4}{\emph{Eur. Phys. J. C}
  {\bfseries 78} (2018) 250}
  [\href{https://arxiv.org/abs/1710.07171}{{\ttfamily 1710.07171}}].

\bibitem{Preskill:1992bf}
J.~Preskill, \emph{{Semilocal defects}},
  \href{https://doi.org/10.1103/PhysRevD.46.4218}{\emph{Phys. Rev. D}
  {\bfseries 46} (1992) 4218}
  [\href{https://arxiv.org/abs/hep-ph/9206216}{{\ttfamily hep-ph/9206216}}].

\bibitem{Sikivie:2006ni}
P.~Sikivie, \emph{{Axion Cosmology}},
  \href{https://doi.org/10.1007/978-3-540-73518-2_2}{\emph{Lect. Notes Phys.}
  {\bfseries 741} (2008) 19}
  [\href{https://arxiv.org/abs/astro-ph/0610440}{{\ttfamily
  astro-ph/0610440}}].

\bibitem{Harigaya:2013vwa}
K.~Harigaya and K.~Mukaida, \emph{{Thermalization after/during Reheating}},
  \href{https://doi.org/10.1007/JHEP05(2014)006}{\emph{JHEP} {\bfseries 05}
  (2014) 006} [\href{https://arxiv.org/abs/1312.3097}{{\ttfamily 1312.3097}}].

\bibitem{Mukaida:2015ria}
K.~Mukaida and M.~Yamada, \emph{{Thermalization Process after Inflation and
  Effective Potential of Scalar Field}},
  \href{https://doi.org/10.1088/1475-7516/2016/02/003}{\emph{JCAP} {\bfseries
  02} (2016) 003} [\href{https://arxiv.org/abs/1506.07661}{{\ttfamily
  1506.07661}}].

\bibitem{Kibble:1982ae}
T.W.B.~Kibble, G.~Lazarides and Q.~Shafi, \emph{{Strings in SO(10)}},
  \href{https://doi.org/10.1016/0370-2693(82)90829-2}{\emph{Phys. Lett. B}
  {\bfseries 113} (1982) 237}.

\bibitem{Vilenkin:1982ks}
A.~Vilenkin and A.E.~Everett, \emph{{Cosmic Strings and Domain Walls in Models
  with Goldstone and PseudoGoldstone Bosons}},
  \href{https://doi.org/10.1103/PhysRevLett.48.1867}{\emph{Phys. Rev. Lett.}
  {\bfseries 48} (1982) 1867}.

\bibitem{Kibble:1982dd}
T.W.B.~Kibble, G.~Lazarides and Q.~Shafi, \emph{{Walls Bounded by Strings}},
  \href{https://doi.org/10.1103/PhysRevD.26.435}{\emph{Phys. Rev. D} {\bfseries
  26} (1982) 435}.

\bibitem{Harigaya:2014waa}
K.~Harigaya, M.~Kawasaki, K.~Mukaida and M.~Yamada, \emph{{Dark Matter
  Production in Late Time Reheating}},
  \href{https://doi.org/10.1103/PhysRevD.89.083532}{\emph{Phys. Rev. D}
  {\bfseries 89} (2014) 083532}
  [\href{https://arxiv.org/abs/1402.2846}{{\ttfamily 1402.2846}}].

\bibitem{Harigaya:2016vda}
K.~Harigaya, T.~Lin and H.K.~Lou, \emph{{GUTzilla Dark Matter}},
  \href{https://doi.org/10.1007/JHEP09(2016)014}{\emph{JHEP} {\bfseries 09}
  (2016) 014} [\href{https://arxiv.org/abs/1606.00923}{{\ttfamily
  1606.00923}}].

\bibitem{Harigaya:2019tzu}
K.~Harigaya, K.~Mukaida and M.~Yamada, \emph{{Dark Matter Production during the
  Thermalization Era}},
  \href{https://doi.org/10.1007/JHEP07(2019)059}{\emph{JHEP} {\bfseries 07}
  (2019) 059} [\href{https://arxiv.org/abs/1901.11027}{{\ttfamily
  1901.11027}}].

\bibitem{Drees:2022vvn}
M.~Drees and B.~Najjari, \emph{{Multi-species thermalization cascade of
  energetic particles in the early universe}},
  \href{https://doi.org/10.1088/1475-7516/2023/08/037}{\emph{JCAP} {\bfseries
  08} (2023) 037} [\href{https://arxiv.org/abs/2205.07741}{{\ttfamily
  2205.07741}}].

\bibitem{Mukaida:2022bbo}
K.~Mukaida and M.~Yamada, \emph{{Cascades of high-energy SM particles in the
  primordial thermal plasma}},
  \href{https://doi.org/10.1007/JHEP10(2022)116}{\emph{JHEP} {\bfseries 10}
  (2022) 116} [\href{https://arxiv.org/abs/2208.11708}{{\ttfamily
  2208.11708}}].

\bibitem{DeLuca:2018mzn}
V.~De~Luca, A.~Mitridate, M.~Redi, J.~Smirnov and A.~Strumia, \emph{{Colored
  Dark Matter}}, \href{https://doi.org/10.1103/PhysRevD.97.115024}{\emph{Phys.
  Rev. D} {\bfseries 97} (2018) 115024}
  [\href{https://arxiv.org/abs/1801.01135}{{\ttfamily 1801.01135}}].

\bibitem{Digman:2019wdm}
M.C.~Digman, C.V.~Cappiello, J.F.~Beacom, C.M.~Hirata and A.H.G.~Peter,
  \emph{{Not as big as a barn: Upper bounds on dark matter-nucleus cross
  sections}}, \href{https://doi.org/10.1103/PhysRevD.100.063013}{\emph{Phys.
  Rev. D} {\bfseries 100} (2019) 063013}
  [\href{https://arxiv.org/abs/1907.10618}{{\ttfamily 1907.10618}}].

\bibitem{Appelquist:2007hu}
T.~Appelquist, G.T.~Fleming and E.T.~Neil, \emph{{Lattice study of the
  conformal window in QCD-like theories}},
  \href{https://doi.org/10.1103/PhysRevLett.100.171607}{\emph{Phys. Rev. Lett.}
  {\bfseries 100} (2008) 171607}
  [\href{https://arxiv.org/abs/0712.0609}{{\ttfamily 0712.0609}}].

\bibitem{Sakurai:1960ju}
J.J.~Sakurai, \emph{{Theory of strong interactions}},
  \href{https://doi.org/10.1016/0003-4916(60)90126-3}{\emph{Annals Phys.}
  {\bfseries 11} (1960) 1}.

\bibitem{Das:1967it}
T.~Das, G.S.~Guralnik, V.S.~Mathur, F.E.~Low and J.E.~Young,
  \emph{{Electromagnetic mass difference of pions}},
  \href{https://doi.org/10.1103/PhysRevLett.18.759}{\emph{Phys. Rev. Lett.}
  {\bfseries 18} (1967) 759}.

\bibitem{Masjuan:2012sk}
P.~Masjuan, E.~Ruiz~Arriola and W.~Broniowski, \emph{{Meson dominance of hadron
  form factors and large-$N_c$ phenomenology}},
  \href{https://doi.org/10.1103/PhysRevD.87.014005}{\emph{Phys. Rev. D}
  {\bfseries 87} (2013) 014005}
  [\href{https://arxiv.org/abs/1210.0760}{{\ttfamily 1210.0760}}].

\bibitem{Witten:1983tw}
E.~Witten, \emph{{Global Aspects of Current Algebra}},
  \href{https://doi.org/10.1016/0550-3213(83)90063-9}{\emph{Nucl. Phys. B}
  {\bfseries 223} (1983) 422}.

\bibitem{Planck:2018vyg}
{\scshape Planck} collaboration, \emph{{Planck 2018 results. VI. Cosmological
  parameters}},
  \href{https://doi.org/10.1051/0004-6361/201833910}{\emph{Astron. Astrophys.}
  {\bfseries 641} (2020) A6}
  [\href{https://arxiv.org/abs/1807.06209}{{\ttfamily 1807.06209}}].

\bibitem{Vachaspati:1991dz}
T.~Vachaspati and A.~Achucarro, \emph{{Semilocal cosmic strings}},
  \href{https://doi.org/10.1103/PhysRevD.44.3067}{\emph{Phys. Rev. D}
  {\bfseries 44} (1991) 3067}.

\bibitem{Hiramatsu:2013qaa}
T.~Hiramatsu, M.~Kawasaki and K.~Saikawa, \emph{{On the estimation of
  gravitational wave spectrum from cosmic domain walls}},
  \href{https://doi.org/10.1088/1475-7516/2014/02/031}{\emph{JCAP} {\bfseries
  02} (2014) 031} [\href{https://arxiv.org/abs/1309.5001}{{\ttfamily
  1309.5001}}].

\bibitem{Kawasaki:2017bqm}
M.~Kawasaki, K.~Kohri, T.~Moroi and Y.~Takaesu, \emph{{Revisiting Big-Bang
  Nucleosynthesis Constraints on Long-Lived Decaying Particles}},
  \href{https://doi.org/10.1103/PhysRevD.97.023502}{\emph{Phys. Rev. D}
  {\bfseries 97} (2018) 023502}
  [\href{https://arxiv.org/abs/1709.01211}{{\ttfamily 1709.01211}}].

\bibitem{Kitajima:2023cek}
N.~Kitajima, J.~Lee, K.~Murai, F.~Takahashi and W.~Yin, \emph{{Gravitational
  waves from domain wall collapse, and application to nanohertz signals with
  QCD-coupled axions}},
  \href{https://doi.org/10.1016/j.physletb.2024.138586}{\emph{Phys. Lett. B}
  {\bfseries 851} (2024) 138586}
  [\href{https://arxiv.org/abs/2306.17146}{{\ttfamily 2306.17146}}].

\bibitem{Turok:1989ai}
N.~Turok, \emph{{Global Texture as the Origin of Cosmic Structure}},
  \href{https://doi.org/10.1103/PhysRevLett.63.2625}{\emph{Phys. Rev. Lett.}
  {\bfseries 63} (1989) 2625}.

\bibitem{Hindmarsh:1991jq}
M.~Hindmarsh, \emph{{Existence and stability of semilocal strings}},
  \href{https://doi.org/10.1103/PhysRevLett.68.1263}{\emph{Phys. Rev. Lett.}
  {\bfseries 68} (1992) 1263}.

\bibitem{Fukuda:2026zoa}
H.~Fukuda, S.~Ge and K.~Harigaya, \emph{{Gravitational Waves and
  Nambu-Goldstone Bosons from Infrared-Finite Cosmic Strings}},
  \href{https://arxiv.org/abs/2610.08254}{{\ttfamily 2610.08254}}.

\bibitem{Kilic:2009mi}
C.~Kilic, T.~Okui and R.~Sundrum, \emph{{Vectorlike Confinement at the LHC}},
  \href{https://doi.org/10.1007/JHEP02(2010)018}{\emph{JHEP} {\bfseries 02}
  (2010) 018} [\href{https://arxiv.org/abs/0906.0577}{{\ttfamily 0906.0577}}].

\bibitem{Beenakker:2024jwh}
W.~Beenakker, C.~Borschensky, M.~Kr{\"a}mer, A.~Kulesza, E.~Laenen,
  J.~Mamu{\v{z}}i{\'c} et~al., \emph{{NNLL-fast 2.0: Coloured sparticle
  production at the LHC run 3 with $\sqrt{S}$ = 13.6 TeV}},
  \href{https://doi.org/10.21468/SciPostPhysCore.7.4.072}{\emph{SciPost Phys.
  Core} {\bfseries 7} (2024) 072}
  [\href{https://arxiv.org/abs/2404.18837}{{\ttfamily 2404.18837}}].

\bibitem{ATLAS:2020syg}
{\scshape ATLAS} collaboration, \emph{{Search for squarks and gluinos in final
  states with jets and missing transverse momentum using 139 fb$^{-1}$ of
  $\sqrt{s}$ =13 TeV $pp$ collision data with the ATLAS detector}},
  \href{https://doi.org/10.1007/JHEP02(2021)143}{\emph{JHEP} {\bfseries 02}
  (2021) 143} [\href{https://arxiv.org/abs/2010.14293}{{\ttfamily
  2010.14293}}].

\bibitem{ATLAS:2021kxv}
{\scshape ATLAS} collaboration, \emph{{Search for new phenomena in events with
  an energetic jet and missing transverse momentum in $pp$ collisions at $\sqrt
  {s}$ =13 TeV with the ATLAS detector}},
  \href{https://doi.org/10.1103/PhysRevD.103.112006}{\emph{Phys. Rev. D}
  {\bfseries 103} (2021) 112006}
  [\href{https://arxiv.org/abs/2102.10874}{{\ttfamily 2102.10874}}].

\bibitem{ATLAS:2024kqk}
{\scshape ATLAS} collaboration, \emph{{A search for R-parity-violating
  supersymmetry in final states containing many jets in pp collisions at $
  \sqrt{s} $ = 13 TeV with the ATLAS detector}},
  \href{https://doi.org/10.1007/JHEP05(2024)003}{\emph{JHEP} {\bfseries 05}
  (2024) 003} [\href{https://arxiv.org/abs/2401.16333}{{\ttfamily
  2401.16333}}].

\bibitem{NNPDF:2014otw}
{\scshape NNPDF} collaboration, \emph{{Parton distributions for the LHC Run
  II}}, \href{https://doi.org/10.1007/JHEP04(2015)040}{\emph{JHEP} {\bfseries
  04} (2015) 040} [\href{https://arxiv.org/abs/1410.8849}{{\ttfamily
  1410.8849}}].

\bibitem{Ahmed:2016otz}
T.~Ahmed, M.~Bonvini, M.C.~Kumar, P.~Mathews, N.~Rana, V.~Ravindran et~al.,
  \emph{{Pseudo-scalar Higgs boson production at N$^3$ LO$_{\text {A}}$ +N$^3$
  LL $'$}}, \href{https://doi.org/10.1140/epjc/s10052-016-4510-1}{\emph{Eur.
  Phys. J. C} {\bfseries 76} (2016) 663}
  [\href{https://arxiv.org/abs/1606.00837}{{\ttfamily 1606.00837}}].

\bibitem{ATLAS:2019fgd}
{\scshape ATLAS} collaboration, \emph{{Search for new resonances in mass
  distributions of jet pairs using 139 fb$^{-1}$ of $pp$ collisions at
  $\sqrt{s}=13$ TeV with the ATLAS detector}},
  \href{https://doi.org/10.1007/JHEP03(2020)145}{\emph{JHEP} {\bfseries 03}
  (2020) 145} [\href{https://arxiv.org/abs/1910.08447}{{\ttfamily
  1910.08447}}].

\bibitem{ATLAS:2025okg}
{\scshape ATLAS} collaboration, \emph{{Search for electroweak-scale dijet
  resonances using trigger-level analysis with the ATLAS detector in
  132{\,}{\,}fb-1 of pp collisions at s=13{\,}{\,}TeV}},
  \href{https://doi.org/10.1103/15p2-bkg8}{\emph{Phys. Rev. D} {\bfseries 112}
  (2025) 092015} [\href{https://arxiv.org/abs/2509.01219}{{\ttfamily
  2509.01219}}].

\bibitem{Pospelov:2007mp}
M.~Pospelov, A.~Ritz and M.B.~Voloshin, \emph{{Secluded WIMP Dark Matter}},
  \href{https://doi.org/10.1016/j.physletb.2008.02.052}{\emph{Phys. Lett. B}
  {\bfseries 662} (2008) 53} [\href{https://arxiv.org/abs/0711.4866}{{\ttfamily
  0711.4866}}].

\bibitem{Kawasaki:2021etm}
M.~Kawasaki, H.~Nakatsuka, K.~Nakayama and T.~Sekiguchi, \emph{{Revisiting CMB
  constraints on dark matter annihilation}},
  \href{https://doi.org/10.1088/1475-7516/2021/12/015}{\emph{JCAP} {\bfseries
  12} (2021) 015} [\href{https://arxiv.org/abs/2105.08334}{{\ttfamily
  2105.08334}}].

\bibitem{Fermi-LAT:2015att}
{\scshape Fermi-LAT} collaboration, \emph{{Searching for Dark Matter
  Annihilation from Milky Way Dwarf Spheroidal Galaxies with Six Years of Fermi
  Large Area Telescope Data}},
  \href{https://doi.org/10.1103/PhysRevLett.115.231301}{\emph{Phys. Rev. Lett.}
  {\bfseries 115} (2015) 231301}
  [\href{https://arxiv.org/abs/1503.02641}{{\ttfamily 1503.02641}}].

\bibitem{Shifman:1978zn}
M.A.~Shifman, A.I.~Vainshtein and V.I.~Zakharov, \emph{{Remarks on Higgs Boson
  Interactions with Nucleons}},
  \href{https://doi.org/10.1016/0370-2693(78)90481-1}{\emph{Phys. Lett. B}
  {\bfseries 78} (1978) 443}.

\bibitem{Abdel-Rehim:2016won}
{\scshape ETM} collaboration, \emph{{Direct Evaluation of the Quark Content of
  Nucleons from Lattice QCD at the Physical Point}},
  \href{https://doi.org/10.1103/PhysRevLett.116.252001}{\emph{Phys. Rev. Lett.}
  {\bfseries 116} (2016) 252001}
  [\href{https://arxiv.org/abs/1601.01624}{{\ttfamily 1601.01624}}].

\bibitem{CTA:2020qlo}
{\scshape CTA} collaboration, \emph{{Sensitivity of the Cherenkov Telescope
  Array to a dark matter signal from the Galactic centre}},
  \href{https://doi.org/10.1088/1475-7516/2021/01/057}{\emph{JCAP} {\bfseries
  01} (2021) 057} [\href{https://arxiv.org/abs/2007.16129}{{\ttfamily
  2007.16129}}].

\bibitem{Manohar:1983md}
A.~Manohar and H.~Georgi, \emph{{Chiral Quarks and the Nonrelativistic Quark
  Model}}, \href{https://doi.org/10.1016/0550-3213(84)90231-1}{\emph{Nucl.
  Phys. B} {\bfseries 234} (1984) 189}.

\bibitem{Georgi:1992dw}
H.~Georgi, \emph{{Generalized dimensional analysis}},
  \href{https://doi.org/10.1016/0370-2693(93)91728-6}{\emph{Phys. Lett. B}
  {\bfseries 298} (1993) 187}
  [\href{https://arxiv.org/abs/hep-ph/9207278}{{\ttfamily hep-ph/9207278}}].

\bibitem{Fukugita:1986hr}
M.~Fukugita and T.~Yanagida, \emph{{Baryogenesis Without Grand Unification}},
  \href{https://doi.org/10.1016/0370-2693(86)91126-3}{\emph{Phys. Lett. B}
  {\bfseries 174} (1986) 45}.

\bibitem{Carrasco-Martinez:2023nit}
J.~Carrasco-Martinez, D.I.~Dunsky, L.J.~Hall and K.~Harigaya,
  \emph{{Leptogenesis in parity solutions to the strong CP problem and Standard
  Model parameters}},
  \href{https://doi.org/10.1007/JHEP06(2024)048}{\emph{JHEP} {\bfseries 06}
  (2024) 048} [\href{https://arxiv.org/abs/2307.15731}{{\ttfamily
  2307.15731}}].

\bibitem{Giudice:2003jh}
G.F.~Giudice, A.~Notari, M.~Raidal, A.~Riotto and A.~Strumia, \emph{{Towards a
  complete theory of thermal leptogenesis in the SM and MSSM}},
  \href{https://doi.org/10.1016/j.nuclphysb.2004.02.019}{\emph{Nucl. Phys. B}
  {\bfseries 685} (2004) 89}
  [\href{https://arxiv.org/abs/hep-ph/0310123}{{\ttfamily hep-ph/0310123}}].

\bibitem{Buchmuller:2004nz}
W.~Buchmuller, P.~Di~Bari and M.~Plumacher, \emph{{Leptogenesis for
  pedestrians}}, \href{https://doi.org/10.1016/j.aop.2004.02.003}{\emph{Annals
  Phys.} {\bfseries 315} (2005) 305}
  [\href{https://arxiv.org/abs/hep-ph/0401240}{{\ttfamily hep-ph/0401240}}].

\bibitem{Covi:1996wh}
L.~Covi, E.~Roulet and F.~Vissani, \emph{{CP violating decays in leptogenesis
  scenarios}}, \href{https://doi.org/10.1016/0370-2693(96)00817-9}{\emph{Phys.
  Lett. B} {\bfseries 384} (1996) 169}
  [\href{https://arxiv.org/abs/hep-ph/9605319}{{\ttfamily hep-ph/9605319}}].

\bibitem{Asaka:1999yd}
T.~Asaka, K.~Hamaguchi, M.~Kawasaki and T.~Yanagida, \emph{{Leptogenesis in
  inflaton decay}},
  \href{https://doi.org/10.1016/S0370-2693(99)01020-5}{\emph{Phys. Lett. B}
  {\bfseries 464} (1999) 12}
  [\href{https://arxiv.org/abs/hep-ph/9906366}{{\ttfamily hep-ph/9906366}}].

\bibitem{Pisarski:1983ms}
R.D.~Pisarski and F.~Wilczek, \emph{{Remarks on the Chiral Phase Transition in
  Chromodynamics}}, \href{https://doi.org/10.1103/PhysRevD.29.338}{\emph{Phys.
  Rev. D} {\bfseries 29} (1984) 338}.

\bibitem{Caprini:2010xv}
C.~Caprini, R.~Durrer and X.~Siemens, \emph{{Detection of gravitational waves
  from the QCD phase transition with pulsar timing arrays}},
  \href{https://doi.org/10.1103/PhysRevD.82.063511}{\emph{Phys. Rev. D}
  {\bfseries 82} (2010) 063511}
  [\href{https://arxiv.org/abs/1007.1218}{{\ttfamily 1007.1218}}].

\bibitem{Schmitz:2020syl}
K.~Schmitz, \emph{{New Sensitivity Curves for Gravitational-Wave Signals from
  Cosmological Phase Transitions}},
  \href{https://doi.org/10.1007/JHEP01(2021)097}{\emph{JHEP} {\bfseries 01}
  (2021) 097} [\href{https://arxiv.org/abs/2002.04615}{{\ttfamily
  2002.04615}}].

\end{thebibliography}\endgroup

\end{document}